\documentclass[journal]{IEEEtran}
\IEEEoverridecommandlockouts
\usepackage{verbatim}%for the comment
\usepackage[utf8x]{inputenc}
\usepackage[pdftex]{graphicx}
\usepackage{epstopdf}
\usepackage{tabularx}
\usepackage{booktabs}
\usepackage[table]{xcolor}
\usepackage{booktabs}
\usepackage{ctable}
\usepackage{multirow}
\usepackage[cmex10]{amsmath}
\usepackage{algorithm}
\usepackage{caption}
\usepackage{subcaption}
\usepackage{algpseudocode}
\usepackage{pifont} 
\usepackage{amssymb}
\usepackage{amsthm}
\usepackage{cite}
\usepackage{textcomp}
\usepackage{array}
\usepackage{enumerate}
\graphicspath{{./Figures/}}

\newtheorem{theorem}{Theorem}
\newtheorem{corollary}{Corollary}

\newcommand{\B}{\operatorname{B}}
\newcommand{\IoT}{\operatorname{IoT}}
\newcommand{\I}{\operatorname{I}}
\newcommand{\PN}{\operatorname{PN}}

\begin{document}
	\bstctlcite{IEEEexample:BSTcontrol}
	
	%---------------------------------------------------------------------------------
	% Title and authors
	%---------------------------------------------------------------------------------
	\title{Spectrum Sharing for Massive Access in Ultra-Narrowband IoT Systems} %The title of the paper
	\author{\IEEEauthorblockN{Ghaith Hattab, \IEEEmembership{Member,~IEEE,} Petar Popovski, \IEEEmembership{Fellow,~IEEE}, and  Danijela Cabric, \IEEEmembership{Senior Member,~IEEE}}
		
		\thanks{An early version of a part of this work was presented in the IEEE International Symposium on Dynamic Spectrum Access Networks (DySPAN), Seoul, South Korea \cite{Hattab2018c}. 
			G. Hattab and D. Cabric are with the Department of Electrical and Computer Engineering, University of California, Los Angeles, CA 90095-1594 USA (email: ghattab@ucla.edu, danijela@ee.ucla.edu). P. Popovski is with the Department of Electronic Systems, Aalborg University, Denmark (email: petarp@es.aau.dk).}
	}

	\maketitle
	
	%---------------------------------------------------------------------------------
	%Abstract
	%---------------------------------------------------------------------------------
	\begin{abstract}
		Ultra-narrowband (UNB) communications has become a signature feature for many emerging low-power wide-area (LPWA) networks. Specifically, using extremely narrowband signals helps the network connect more Internet-of-things (IoT) devices within a given band. It also improves robustness to interference, extending the coverage of the network. In this paper, we study the coexistence capability of UNB networks and their scalability to enable massive access. To this end, we develop a stochastic geometry framework to analyze and model UNB networks on a large scale. The framework captures the unique characteristics of UNB communications, including the asynchronous time-frequency access, signal repetition, and the absence of base station (BS) association. Closed-form expressions of the transmission success probability and network connection density are presented for several UNB protocols. We further discuss multiband access for UNB networks, proposing a low-complexity protocol. Our analysis reveals several insights on the geographical diversity achieved when devices do not connect to a single BS, the optimal number of signal repetitions, and how to utilize multiple bands without increasing the complexity of BSs. Simulation results are provided to validate the analysis, and they show that UNB communications enables a single BS to connect thousands of devices even when the spectrum is shared with other networks.
	\end{abstract}

	%---------------------------------------------------------------------------------
	%Index Words
	%---------------------------------------------------------------------------------

	%---------------------------------------------------------------------------------
	%Section: Introduction
	%---------------------------------------------------------------------------------
	\section{Introduction} \label{sec:UNB}
	
	The emergence of massive Internet-of-things (IoT) applications has spurred the development of a new class of networks, known as low-power wide-area (LPWA) networks \cite{Raza2017,CentenaroZorzi2016,Xiong2015}. The majority of LPWA networks rely on using the unlicensed spectrum due to its low capital expenditure; they provide long-range connectivity, primarily using bands sub-1GHz due to their favorable propagation conditions, and they use lightweight access protocols to limit the communication overhead and extend end-devices' lifetime \cite{Raza2017}. One popular variant of LPWA networks, which is the focus of this paper, is the ultra-narrowband (UNB) network.
	
	In UNB networks, data communications is done using extremely narrowband signals to connect a large number of devices, addressing the \emph{intra-network} sharing problem, i.e., how a large number of IoT devices share the same band. Equally important, concentrating the signal energy into ultra-narrow channels improves robustness to interference from other technologies using the same unlicensed spectrum, addressing the \emph{inter-network sharing} problem. Indeed, the uplink (UL) signal bandwidth is a few hundred Hertz. In addition, these networks rely on simple asynchronous, or ALOHA-like, access protocols, and more interestingly, IoT devices avoid prior association and synchronization with any UNB base station (BS); in essence, IoT devices usually operate in a broadcast mode, transmitting their packets at any time and frequency. The primary restrictions on these devices are related to how many packets per day each can send and the bandwidth of the band that comprises the different channels a device can pick from. Furthermore, to combat the absence of network acknowledgments and the presence of interfering incumbent networks, signal repetition is used; in particular, each IoT packet is sent multiple times, one after another, each at a different frequency within the predetermined multiplexing band. 
	
	%-----------------------------------------------------------
	%Subsection: Related work
	%-----------------------------------------------------------
	\subsection{Related Work}\label{sec:UNB_related}
	
\begin{figure}[t!]
		\center
		\scriptsize
		\includegraphics[width=0.5\textwidth]{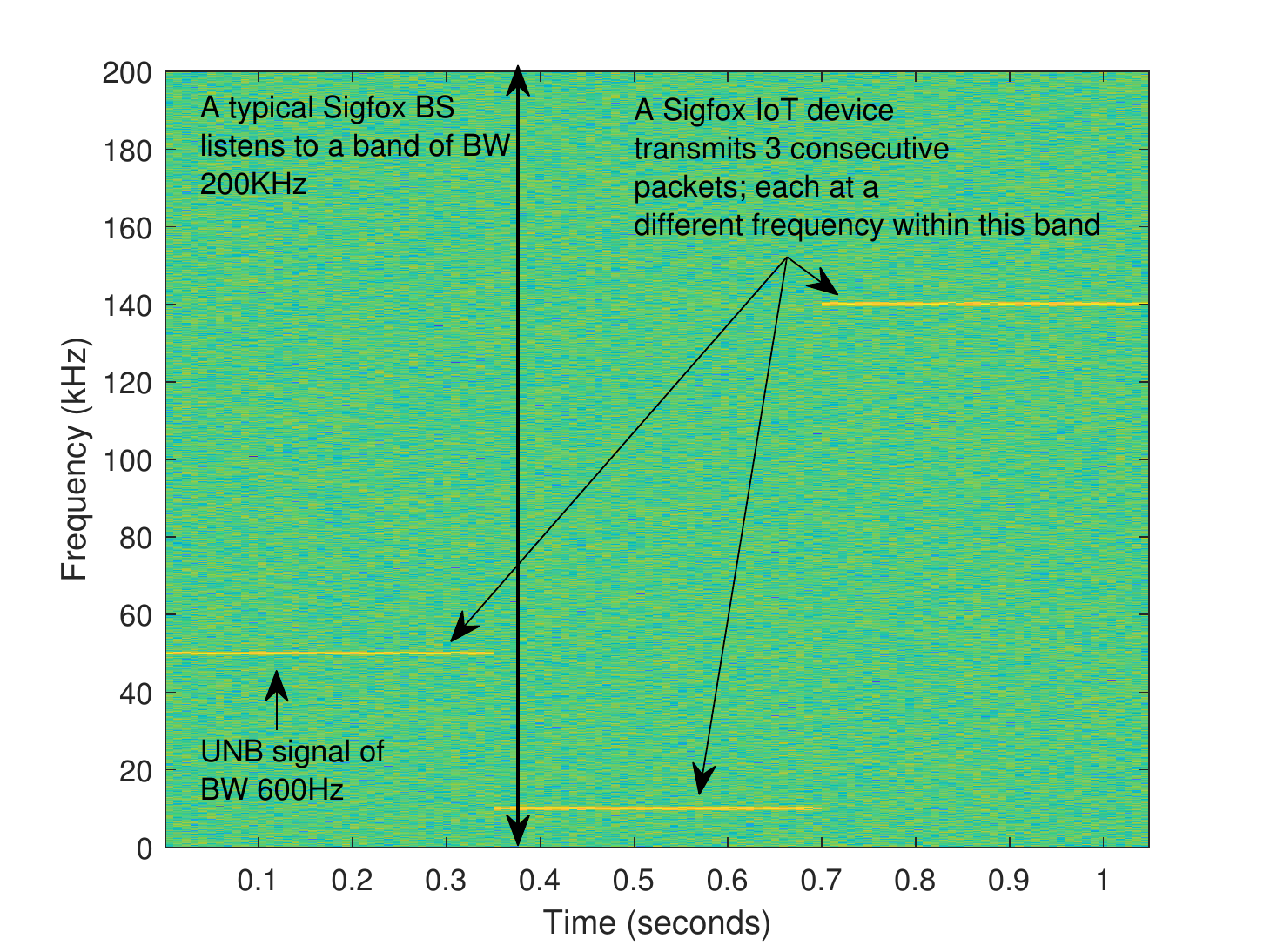}
		\caption{Spectrogram of one Sigfox packet of bandwidth 600Hz. The packet is sent three times at randomly selected frequencies within a multiplexing band of bandwidth 200KHz.}
		\label{fig:SigfoxSpectogram}
	\end{figure}

	A list of few IoT networks that use proprietary UNB technologies is given in Table \ref{tab:UNB}, with Sigfox being one of the most popular LPWA networks \cite{Sigfox2017,Sigfox2019}. Fig. \ref{fig:SigfoxSpectogram} shows the spectrogram of one Sigfox packet, which is sent three times randomly hopping from one frequency to another. In this paper, our objective is to model, analyze, and optimize these UNB networks, considering their unique communications characteristics and examining their scalability, in terms of the connection density of IoT devices, and their robustness to interference from incumbent networks. 
	
	Network-level analytical frameworks for UNB communications remain largely unexplored. Indeed, and to the best of our knowledge, prior work captures one or few of UNB communication properties. For instance, early works have considered single cells and ignored the presence of incumbents \cite{Goursaud2016,Mo2016,Mo2016a}. The more recent work in \cite{Mo2018} has considered the performance of UNB networks when two BSs are used to decode IoT packets, although signal repetition is ignored. These works assume a specific geometry of the network, whereas in this work we use stochastic geometry \cite{Haenggi2012} to help derive fundamental limits and performance trends of UNB networks that are not restricted to particular deployment geometries. While stochastic geometry has become a popular tool to model cellular networks \cite{ElSawyWin2017}, it is also useful for UNB networks as it captures their spatial randomness and the fact that UNB devices transmit at random times and frequencies.
	
	\begin{table}[!t]
		\caption{Examples of UNB technologies}
		\label{tab:UNB}
		\centering
		\small 
		\begin{tabular}{|l|c|}
			\hline
			\textbf{UNB Technology }  				&  \textbf{Bandwidth (Hz)}\\\hline
			\multirow{1}{*}{Sigfox \cite{Sigfox2019}  } & 600 (US) and 100 (Europe)\\	\hline
			WavIoT NB-Fi \cite{WAVIoT2016}   & 100\\\hline
			NWave Weightless-N \cite{Weightless2018} &200\\\hline
			Telensa \cite{Telensa2018} &500\\\hline 
		\end{tabular}
	\end{table}
	
	In addition, the aforementioned works have quantified the UNB network performance in terms of the probability of collision, i.e., two or more devices picking the same time-frequency resources. This metric is pessimistic and, in fact, becomes inaccurate when UNB networks share the spectrum with devices that use wider channels. For example, other LPWA networks, such as LoRa, share the same spectrum with UNB networks. In this case, a LoRa IoT device can transmit signals at a bandwidth of 250KHz, completely covering the 200KHz multiplexing band used by the Sigfox network. Thus, under the aforementioned metric, UNB communications is impossible when the LoRa device is active! 
	In this work, we focus on the received signal-to-noise-plus-interference ratio (SINR) of UNB packets, i.e., a packet is successfully decoded if the SINR is above some threshold. Such SINR analysis of IoT communications has been considered in \cite{Lim2018, Jiang2018a,Centenaro2017,AlcarazLopez2018,Hattab2018b}, but these works consider LoRa or cellular networks, where access is vastly different compared to UNB communications. Indeed, a cellular narrowband-IoT device connects to a specific cell and communicates with its associated BS over a specific channel. From a modeling perspective, cell association divides the region into Voronoi cells, affecting the interference characteristics \cite{Jiang2018a}. For example, the interference seen at a BS in a time-frequency resource comes from devices located outside the cell in synchronous access, whereas it only stems from multiple devices using the same preamble in grant-free access. In contrast, we consider a cell-free network, i.e., interference can come from any active device, and with asynchronous access, it can come from devices overlapping in time, frequency, or both. Equally important, in our model, the packet can be decoded at any BS in the vicinity of the device, thus offering additional diversity compared to the cellular model that only considers a single BS to decode data packets. Similarly, in LoRa networks, spread spectrum signals are used instead of UNB signals, and thus the interference across LoRa devices depends on the spreading factor used.

	%-----------------------------------------------------------
	%Subsection: Contributions
	%-----------------------------------------------------------
	\subsection{Contributions}\label{sec:UNB_contributions}
	
	\begin{figure*}[t!]
		\center
		\includegraphics[width=0.9\textwidth]{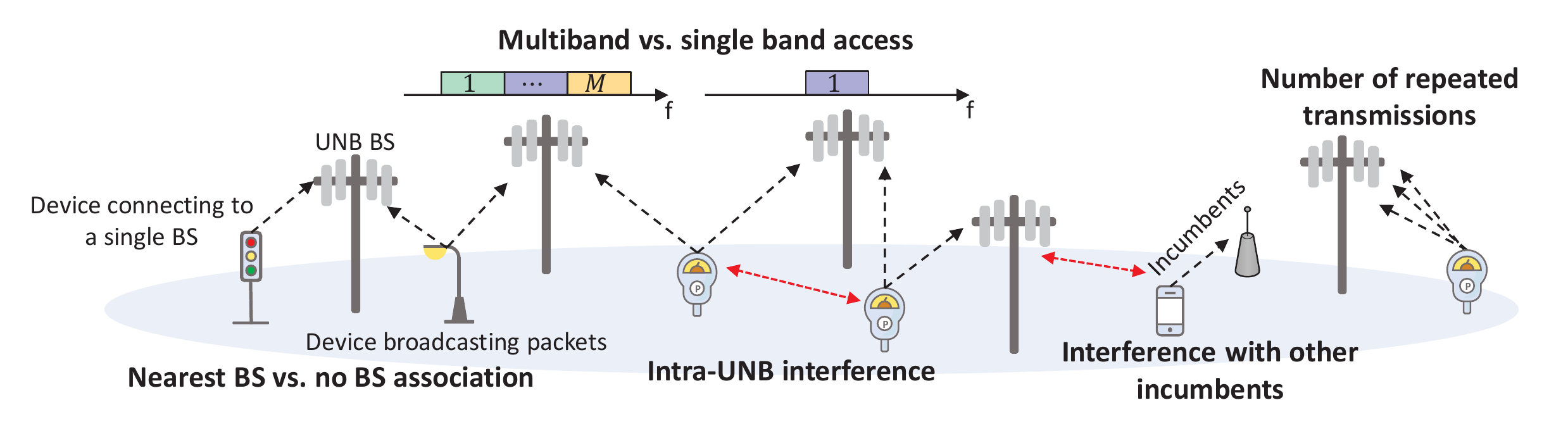}
		\caption{The stochastic framework considers several factors to model and analyze UNB networks.}
		\label{fig:UNB_system}
	\end{figure*}
	
	The main contributions of this work are twofold. First, we develop a stochastic geometry framework to model UNB networks that share the spectrum with other networks. The framework captures the unique characteristics of UNB communications and considers several factors, as shown in Fig. \ref{fig:UNB_system}, including the number of repetitions, the bandwidth utilized by the network, and the density of BSs listening to a particular packet. To this end, we derive closed-form expressions for the transmission success probability, which is defined as the probability of having at least one of the packet's transmissions successfully decoded by the network. The obtained expressions are then used to determine the transmission capacity of UNB networks \cite{Weber2012}, i.e, the maximum density of IoT devices that can be reliably covered in the network for a given density of UNB BSs. The analysis identifies the key parameters that affect the success probability and transmission capacity, gleaning insights on the geographical diversity achieved when the IoT device does not associate with a single BS as well as the optimal number of repetitions.
	
	Second, our analysis reveals the interplay between space and frequency, i.e., the density of UNB BSs and the number of bands used by the network when each BS is equipped with a high-complexity multiband receiver. Specifically, to meet a target success probability performance, the UNB network can offer additional frequency bands to help IoT devices avoid incumbents or can deploy more BSs, in case of single-band access, to increase the geographical diversity. Such insight motivates us to propose a low-complexity multiband access protocol, where the network uses multiple bands, yet each BS listens to a single one. We show how such a low-complexity protocol can improve the network's performance compared to existing single-band protocols.
	
	Simulation results are presented to validate the theoretical expressions. It is shown that a single UNB BS can connect thousands of devices even in the presence of interfering technologies. Thus, UNB communications is not only a potential paradigm for LPWA networks but also for massive access in beyond-5G wireless networks. Specifically, with the emergence of NB-IoT \cite{Xu2018}, NR-Unlicensed \cite{Qualcomm2019}, and grant-free access \cite{Liu2018a}, future cellular networks can provide UNB-like protocols, exploring the possibility of asynchronous time-frequency-space access for delay-tolerant massive IoT traffic.
	
	%-----------------------------------------------------------
	%Section: System model
	%-----------------------------------------------------------
	\section{System Model}\label{sec:UNB_model}
	
	%--------------------------------
	%Subsection: UNB network topology and transmission model
	%--------------------------------
	\subsection{UNB network topology and transmission model}
	
	We consider a random spatial topology of BSs and UNB IoT devices. Specifically, for the UNB network, we assume that BSs' locations and IoT devices are generated from homogeneous Poisson Point processes (HPPP) $\Phi_{\B}$ and $\Phi_{\IoT}$, with densities $\lambda_{\B}$ and $\lambda_{\IoT}$, respectively. 
	
	The UNB IoT device transmits signals at power $P_{\IoT}$, occupying a bandwidth $b$. Each signal is sent $N$ times, consecutively over time, yet hopping from one frequency to another. These signals are commonly sent within a predetermined \emph{multiplexing band} of bandwidth $B$. In this work, we generalize the analysis to multiband access, i.e., we assume there are $M$ multiplexing bands, each of bandwidth $B$. For the temporal generation of IoT traffic, UNB devices randomly send packets over time at the following rate
	\begin{equation}
	\lambda_{\operatorname{T}} = \frac{K\times T}{T_{\operatorname{tot}}},
	\end{equation}
	where $K$ is the number of unique reports or packets that can be transmitted over a total duration of $T_{\operatorname{tot}}$, and $T$ is the duration of each transmission. For example, in Sigfox \cite{Sigfox2019}, the device can send up to $K=6$ packets per hour, i.e., $T_{\operatorname{tot}}=3600$s. Each Sigfox packet is typically 26 bytes, and thus at $b=600$Hz, the transmission duration is $T = \frac{26\times 8}{600}\approx 0.35$s.\footnote{We consider all UNB devices to have the same payload. However, the analysis in this work can be generalized to consider a multi-tier UNB network, where different tiers can be characterized by different payloads.} We note that the limits on $K$ and $T$ are due to the sub-1GHz regulations on duty cycles and dwell times.
	
	Since the transmissions are sent over extremely narrowband channels, it is difficult to send a signal at a specific channel and carrier frequency. In fact, the frequency offset in a low-cost oscillator could be higher than the signal bandwidth, making it impractical to assume a perfectly channelized system for UNB networks. For this reason, UNB networks are commonly assumed to have an unslotted frequency access \cite{Do2014}. In particular, let the $n$-th transmission be sent at carrier frequency $f_n$; then, in unslotted frequency access, we have
	$
	f_n\sim\mathcal{U}\left(f_c - \frac{M\cdot B + b}{2},f_c + \frac{M\cdot B - b}{2}	\right)$, 
	where $f_c$ is the center frequency of the spectrum used by the UNB network and $\mathcal{U}(u_1,u_2)$ is the uniform distribution on the interval $[u_1,u_2]$. While UNB networks are actually unslotted, we generalize the analysis to consider the slotted case in time and/or frequency. 
	
\begin{figure}[t!]
		\center
		\includegraphics[width=0.5\textwidth]{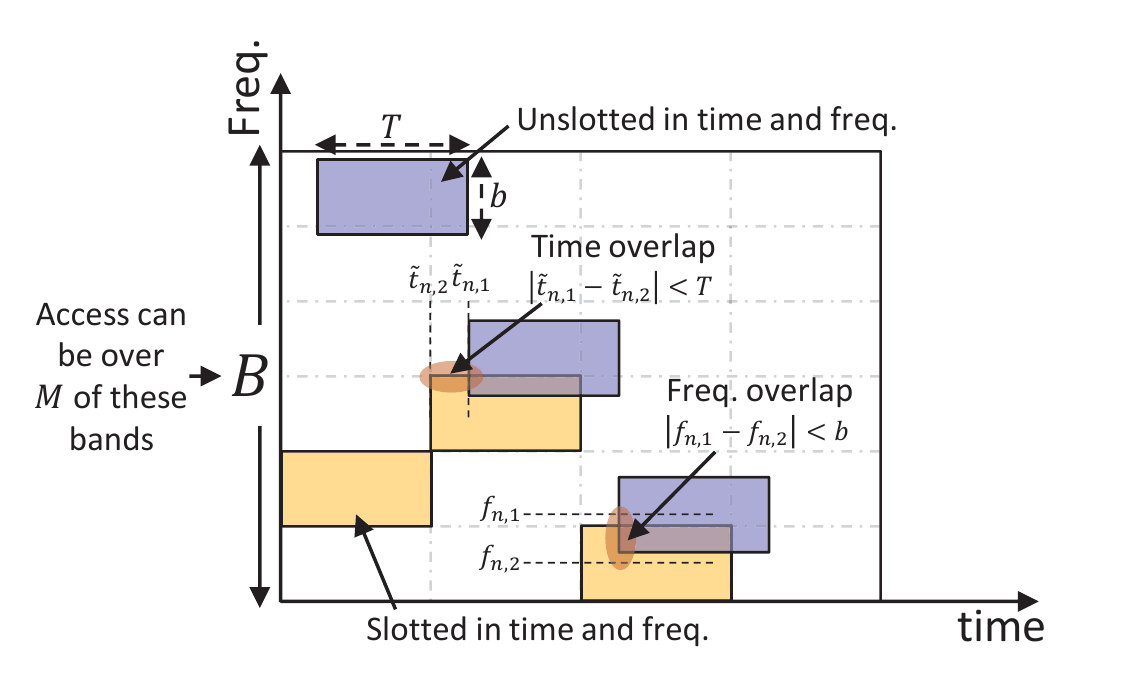}
		\caption{Access can be slotted or unslotted in time and/or frequency.}
		\label{fig:UNB_accessScenarios}
	\end{figure}

	In the slotted case, a device interferes with another if both devices are transmitting over the same slot, which occurs with probability $T/T_{\operatorname{tot}}$ and $b/(M\cdot B)$ in time and frequency, respectively. In the unslotted time (or frequency) case, interference between two devices occurs if their signals have any overlap in time (or frequency). For instance, if $\tilde t_{n,1}$ and $\tilde t_{n,2}$ are the start transmission times of device one and two, then an interference occurs if $|\tilde t_{n,1}-\tilde t_{n,2}|<T$, and thus two devices interfere with probability $2 T/T_{\operatorname{tot}}$. Similarly, if  $f_{n,1}$ and $f_{n,2}$ are the center frequencies of the two devices, then an interference occurs if $|f_{n,1} - f_{n,2}|<b$, which happens with probability $2b/(M\cdot B)$. In other words, an unslotted system in time (or frequency) doubles the probability of a device interfering with another one in comparison with a slotted system in time (or frequency). Similarly, a completely asynchronous time-frequency system quadruples the probability of interference compared to a time-frequency slotted system. For a compact notation, we use $1\leq \beta_{T}\leq 2$, where $\beta_{T}=1$ and $\beta_{T}=2$ denote a slotted and an unslotted system in time, respectively. Note that $\beta_T\in(1,2)$ can denote different levels of time synchronization or interference overlap tolerance. An identical notation is used for frequency, where we use $\beta_F$. Fig. \ref{fig:UNB_accessScenarios} illustrates the different access scenarios.
	
	%--------------------------------
	%Subsection: Interfering networks
	%--------------------------------
	\subsection{Interfering networks}
	
	For interfering or incumbent networks that share the spectrum with the UNB network, we consider two different types. In particular, Type-I comprises a single incumbent network, where interfering devices are generated from another independent HPPP $\Phi_{\I,0}$ with density $\lambda_{\I,0}$. Each interfering device transmits signals at power $P_{\I}$ over a bandwidth $B_{\I,0}\gg b$, where this bandwidth is assumed to be overlapped with the spectrum used by the UNB network and $B_{\I,0}>B$ is feasible. The interfering network could be a WiFi network or another IoT-based network, e.g., LoRa \cite{CentenaroZorzi2016,Bor2016}.
	
	In Type-II model, we assume each multiplexing band is occupied by possibly a different interfering network, where the $m$-th band is occupied by interfering devices that are generated from the independent HPPP $\Phi_{\I,m}$ with density $\lambda_{\I,m}$. Each device transmits signals at power $P_{\I}$ over a bandwidth $b \ll B_{\I,m}\leq B$.  
	To summarize, the set of interferes can be given as\footnote{While it is straightforward to generalize the analysis with different transmit powers across incumbent networks, we assume the same power for easier exposition, i.e., $P_{\I,m} = P_{\I,0} = P_{\I}$. In addition, the analysis in this work naturally extends to include a composition of the two types of interfering networks.}
	\begin{equation}
	\bar \Phi_{\I} = \left\{
	\begin{array}{ll}
	\Phi_{\I,0}, & \text{Type-I}\\
	\bigcup\limits_{m=1}^M \Phi_{\I,m}, & \text{Type-II}\\
	\end{array}\right..
	\end{equation}
	
	%--------------------------------
	%Subsection: SINR expression
	%--------------------------------
	\subsection{SINR expression}
	
	Consider a typical UNB device at a distance $x_{j}$ from the $j$-th BS. Then, the SINR of the $n$-th message at this BS can be expressed as
	
	\begin{equation}
	\begin{aligned}
	\label{eq:SINR}
	\operatorname{SINR}_{n,j} = \frac{h_n x_j ^{-\alpha} }{\hat P_N +\underset{I_{\operatorname{UNB}}}{\underbrace{\sum_{u\in \tilde \Phi_{\IoT, n}} f_u y_{u,j}^{-\alpha} }} + \underset{I_{\operatorname{INC}}}{\underbrace{\sum_{k\in\tilde \Phi_{\I, n}} \hat P_{\I} f_k y_{k,j}^{-\alpha}}}},
	\end{aligned} 
	\end{equation}
	
	where $\alpha$ is the path loss exponent, $h$ and $f$ are exponentially distributed with unit power, $\hat P_{\I}= \frac{P_{\I}\cdot b/B_{\I}}{P_{\IoT}}$, $\hat P_N= \frac{P_N}{P_{\IoT}}$, $P_N$ is the noise power, $y_{u,j}$ and $y_{k,j}$ are the distances from the $j$-th BS to an interfering UNB device and an interfering incumbent, respectively, and $\tilde \Phi_{\IoT, n}$ and $\tilde \Phi_{\I, n}$ are the set of interfering UNB devices and incumbents during the $n$-th transmission, respectively.

	%--------------------------------
	%Subsection: Performance metrics
	%--------------------------------
	\subsection{Performance metrics}
	
	Different UNB protocols are compared in terms of \emph{the transmission success probability}, $\mathbb{P}_s$; it is defined as the probability that at least one of the $N$ messages is decoded successfully, i.e., the signal received SINR is above a threshold $\tau$. Further, the success probability can be computed using the complementary cumulative distribution function (CDF) of the maximum SINR among the $N$ messages. This expression is useful to determine \emph{the transmission capacity} or \emph{connection density} of the network, which is defined as the maximum number of devices that can be supported for a given density of BSs and success probability constraint $\gamma\in(0,1)$. More formally, the success probability can be written as a function of the density of IoT devices, i.e., $\mathbb{P}_s = F(\lambda_{\IoT})$; thus, the transmission capacity is defined as $\mathbb{C}(\gamma) = \gamma \cdot F^{-1}(\gamma)$ \cite{Weber2012}.

	%-----------------------------------------------------------
	%Section: Performance Analysis of UNB Protocols
	%-----------------------------------------------------------
	\section{Performance Analysis of UNB Protocols}\label{sec:UNB_analysis} 
	
	In this section, we analyze the performance of two variants of UNB protocols: in the first one, the IoT device only communicates with the nearest BS. In the second one, the IoT device operates in a broadcast mode, i.e., there is no BS association. While the latter is adopted by Sigfox, analyzing the former helps highlight the impact of the geographical diversity achieved when the device broadcasts its packets instead of sending them to one BS. Unless otherwise stated, we consider random frequency hopping for repeated transmissions, whereas we study pseudorandom hopping in Section \ref{sec:PNhopping}.	
	
	%--------------------------------
	%Subsection: Performance with nearest BS association
	%--------------------------------
	\subsection{Performance with nearest BS association} 
	
	Let the device be connected to the nearest BS $j$. Then, a transmission is considered successful if this particular BS decodes at least one of the $N$ messages, i.e., 
	\begin{equation}
	\label{eq:PsGeneralNearestBS}
	\begin{aligned}
	\mathbb{P}_s^{\operatorname{nearest}}  &= 1 - \text{Pr}\{\text{No message is decoded at the nearest BS}\}.
	\end{aligned}
	\end{equation}
	Clearly, a BS cannot decode any of the $N$ messages if the message with the maximum SINR is below the decoding threshold $\tau$. More formally, the $j$-th BS fails to decode the $N$ messages with probability 
	\begin{equation}
	\label{eq:failProb}
	\mathbb{Q}_j = \text{Pr}\left( \operatorname*{max}_{n\in\{1,2,\cdots,N\}} \operatorname{SINR}_{n,j} \leq \tau \right),
	\end{equation}
	which means that the success probability in (\ref{eq:PsGeneralNearestBS}) can be rewritten as
	\begin{equation}
	\label{eq:PsGeneralNearestAssociation}
	\mathbb{P}_s^{\operatorname{nearest}}   = 1 - \mathbb{E}_{x} \left[\mathbb{Q}_j\right],
	\end{equation}
	where the expectation is with respect to the location of the nearest BS in $\Phi_{\B}$. 
	We can simplify (\ref{eq:failProb}) as follows
	\begin{equation}
	\label{eq:failProbR}
	\begin{aligned}
	\mathbb{Q}_j &= \text{Pr}\left(\operatorname{SINR}_{1,j} \leq \tau, \operatorname{SINR}_{2,j} \leq \tau, \cdots, \operatorname{SINR}_{N,j}\leq \tau \right)\\
	&\stackrel{(a)}{=}  \mathbb{E}_{\tilde \Phi_{\IoT},\tilde\Phi_{\I},f_u,f_k}\textstyle \left[\text{Pr}\textstyle\big(h_n \leq \tau x_j^\alpha (\hat P_N +I_{\operatorname{UNB}} + I_{\operatorname{INC}})\big)^N\right]\\
	&\stackrel{(b)}{=} \left(\mathbb{E}_{\tilde \Phi_{\IoT},\tilde\Phi_{\I},f_u,f_k} \left[(1-e^{-\tau x_j^\alpha (\hat P_N +I_{\operatorname{UNB}} + I_{\operatorname{INC}})})\right]\right)^N,
	\end{aligned}
	\end{equation}
	where $(a)$ follows as each signal is sent over a different channel with independent fading\footnote{Channels can be assumed to be independent even if frequency hopping is done over a narrowband of bandwidth $B$ because the duration of each UNB signal is long enough to justify different channels across messages \cite{Sigfox2017}.} and $(b)$ follows from the CDF of $h_n$ and the fact that the set of interfering UNB devices across transmissions is an independent thinned HPPP since UNB devices randomly transmit over time and frequency. Note that an assumption here, that we make throughout this paper, is that each IoT transmission will experience an independent set of interfering incumbents, irrespective of the interfering network model used. This assumption simplifies the analysis and is reasonable as we assume incumbent transmitters have shorter transmission durations (recall $B_{\I}\gg b$). For instance, Sigfox transmission duration is $T\approx0.35$s in the US (or $2$s in Europe), whereas other incumbents, e.g. LoRa, typically have durations of the order of few tens of milliseconds for similar packet sizes \cite{Sornin2017}. 
	
	To obtain a closed-form expression of (\ref{eq:failProbR}), we need to determine the spatial process of interfering UNB devices, i.e., $\tilde \Phi_{\IoT,n}$, and interfering incumbent devices, i.e., $\tilde \Phi_{\I,n}$. Since UNB devices randomly pick time-frequency slots, the set of interfering UNB devices $\tilde \Phi_{\IoT,n }$ is essentially a thinned $\Phi_{\IoT}$, which is still an HPPP with the same density across all transmissions \cite{Haenggi2012}. The density of this thinned process is 
	\begin{equation}
	\label{eq:interferenceIoTDensity}
	\tilde{\lambda}_{\IoT} = N\cdot \beta_{T}\lambda_{\operatorname{T}} \cdot \frac{\beta_{F}b}{M\cdot B}\cdot \lambda_{\IoT}.
	\end{equation}
	This is derived as follows. The density of IoT devices is $\lambda_{\IoT}$, and the portion of these devices that would have overlapping transmissions in time is $N \beta_{T} \lambda_{\operatorname{T}}$. However, not all of those devices will have the same carrier frequency as the typical device. Indeed, the probability of a device's signal overlapping in frequency with that of the typical one is $\frac{\beta_{F}b}{M\cdot B}$, for which (\ref{eq:interferenceIoTDensity}) follows.
	
	In a similar manner, the set of interfering incumbents is another HPPP process with density 
	\begin{equation}
	\tilde{\lambda}_{\I} = \left\{
	\begin{array}{ll}
	\min\{1,\frac{B_{\I,0}}{M\cdot B}\}\lambda_{\I,0}, &\text{Type-I}\\
	\frac{1}{M}\sum_{m=1}^M\frac{B_{\I,m}}{B}\lambda_{\I,m}, &\text{Type-II}\\
	\end{array}\right..
	\end{equation}
	This follows for Type-I since there is a single network that can occupy a BW of $B_{\I,0}$ in a spectrum of BW $M\cdot B$. Similarly, for Type-II, each incumbent network occupies a bandwidth $B_{\I,m}$ within the band, and each band is selected with a probability of $1/M$ by the typical UNB device.
	The next theorem provides the success probability of UNB access with nearest BS association.
	
	\begin{theorem}\label{th:Ps_nearest}
		In an interference-limited network, where $\hat P_N \rightarrow 0$, the success probability of UNB access under nearest BS association is given as
		\begin{equation}
		\label{eq:Ps_nearest}
		\mathbb{P}_s^{\operatorname{nearest}} = 1-\textstyle \sum_{k=0}^N \binom{N}{k} (-1)^k \left(1+k\xi^{-1} \tau^\delta\frac{\tilde{\lambda}_{\IoT}+\hat P_{\I}^\delta \tilde{\lambda}_{\I}}{\lambda_{\B}}\right)^{-1},
		\end{equation}
		where $\delta=2/\alpha$ and $\xi= \frac{ \sin(\pi\delta)}{\delta\pi}$.
	\end{theorem}
	
	\noindent \emph{Proof}: See Appendix \ref{app:Ps}. $\hfill\blacksquare$
	
	The transmission capacity, in this case, can be computed numerically when $N>1$. When $N=1$, we have the following result.
	\begin{corollary}
		The transmission capacity of UNB access under nearest BS and $N=1$ is
		\begin{equation}
		\label{eq:TC_nearest}
		{\mathbb{C}}^{\operatorname{nearest}}(\gamma) = \frac{\gamma M\cdot B}{\beta_{T}\beta_{F} b \lambda_{\operatorname{T}}}\left(\frac{\xi \tau^{-\delta} \lambda_{\B}}{(\frac{\gamma}{1-\gamma})}-\hat P_{\I}^\delta\tilde \lambda_{\I}\right).
		\end{equation}	 $\hfill\blacksquare$
	\end{corollary}
	\noindent  We have the following key observations. First, the presence of an interfering network reduces the transmission capacity, as the expression in parentheses decreases with $\tilde \lambda_{\I}$. Second, increasing the number of bands can bring two gains: it can reduce the effective density of interferers, particularly for Type-I incumbents, and it scales the UNB transmission capacity, i.e., the term outside the parentheses is also increased. Note that for $M=1$ and Type-I interfering networks, we get the performance of UNB networks with single-band access. Third, as expected, the transmission capacity depends on the device's requirements, i.e., number of packets, packet size, etc., which are all captured in $\lambda_{\operatorname{T}}$. Finally, from $\beta_{T}$ and $\beta_{F}$ in (\ref{eq:TC_nearest}), it is observed that a completely synchronous system, i.e. $\beta_{T}=\beta_{F}=1$, quadruples the transmission capacity compared to an asynchronous time-frequency system, i.e., $\beta_{T}=\beta_{F}=2$ .
	
	%--------------------------------
	%Subsection: Performance with no BS association
	%--------------------------------
	\subsection{Performance with no BS association} 
	
	In this section, we consider a UNB network similar to Sigfox, where a device operates in a broadcast mode; hence, the transmission is successful if \emph{at least one} BS decodes at least one of the $N$ messages. In this case, the success probability is 
	\begin{equation}
	\begin{aligned}
	\label{eq:PsGeneralNoAssociation}
	\mathbb{P}_s  &= 1 - \text{Pr}\{\text{No message is successfully decoded at any BS}\}\\
	&= 1 - \mathbb{E}_{\Phi_{\B}} \left[\prod_{b\in\Phi_{\B}} \mathbb{Q}_b\right].
	\end{aligned}
	\end{equation}
	Compared to (\ref{eq:PsGeneralNearestAssociation}), the expectation here is with respect to the set of UNB BSs instead of just the distance to the nearest one. The success probability performance is given in the next theorem. 
	
	\begin{theorem}\label{th:Ps}
		In an interference-limited network, the success probability of UNB access under no BS association is given as 
		\begin{equation}
		\label{eq:Ps}
		\mathbb{P}_s = 1 - \exp\left(-\xi \tau^{-\delta} H_{N}\cdot \frac{\lambda_{\B}}{\tilde{\lambda}_{\IoT}+\hat P_{\I}^\delta\tilde{\lambda}_{\I}}\right),
		\end{equation}
		where $H_N$ is the harmonic number, i.e., $H_N=-\sum_{k=1}^N \binom{N}{k} (-1)^k k^{-1}$. 
	\end{theorem}
	\noindent \emph{Proof}: See Appendix \ref{app:Ps}. $\hfill\blacksquare$

	Using Theorem \ref{th:Ps}, we can obtain the transmission capacity in closed-form as follows. 
	\begin{corollary}
		The transmission capacity of the UNB network with no BS association is 
		\begin{equation}
		\label{eq:TC}
		\mathbb{C}(\gamma) = \frac{\gamma M B}{\beta_{T}\beta_{F} b \lambda_{\operatorname{T}}}\left(\frac{\xi \tau^{-\delta} H_N\lambda_{\B}}{N \ln(\frac{1}{1-\gamma})}-\frac{\hat P_{\I}^\delta\tilde \lambda_{\I}}{N}\right).
		\end{equation}	 $\hfill\blacksquare$
	\end{corollary}
	Next, we closely examine the performance gains of operating in a broadcast mode instead of connecting to a single BS.
	
	%--------------------------------
	%Subsection: The role of geographical diversity
	%--------------------------------
	\subsection{The role of geographical diversity} 
	
	In geographical diversity, also known as macro diversity, multiple physically apart BSs are used to receive the same signal. Thus, the signal can experience multiple independently fading channels, substantially improving transmission reliability. We distinguish this diversity method from spatial diversity, which is used by having multiple antennas at the same BS. The latter, in essence, affects the distribution of $h$ and $f$ in (\ref{eq:SINR}), whereas geographical diversity affects the distances.  
	
	In what follows, we assume a single-transmission per packet, i.e., $N=1$, to factor out the impact of repetition diversity. Let the target success probability be $\mathbb{P}_s \geq \epsilon$ and $B_{\I,0} \leq M\cdot B$. Then, we have the following corollary. 
	\begin{corollary}\label{corollary:geographicalDiversity}
		To achieve a success probability of at least $\epsilon$, the following constraint  must be satisfied 
		\begin{equation}
		\label{eq:UNB_lowerBound_Success}
		M\lambda_{\B} \geq c g(\epsilon),
		\end{equation}	
		where $c= \left(\beta_{T}\lambda_{\operatorname{T}} \beta_{F} \frac{b}{B} \lambda_{\IoT} + \hat P_{\I}^\delta \tilde{\tilde{\lambda}}_{\I}\right)\xi^{-1} \tau^{\delta}$,
		\begin{equation}
		\tilde{\tilde{\lambda}}_{\I} = \left\{
		\begin{array}{ll}
		\frac{B_{\I,0}}{B}\lambda_{\I,0}, &\text{Type-I}\\
		\frac{1}{B}\sum_{m=1}^MB_{\I,m}\lambda_{\I,m}, &\text{Type-II}\\
		\end{array}\right., 
		\end{equation}
		and 
		\begin{equation}
		g(\epsilon) = \left\{
		\begin{array}{ll}
		\frac{\epsilon}{1-\epsilon},& \text{Nearest BS}\\
		\ln\left(\frac{1}{1-\epsilon}\right),& \text{No BS association}
		\end{array}\right..
		\end{equation}
	\end{corollary}
	\noindent \emph{Proof}: For $N=1$, we can rewrite (\ref{eq:Ps_nearest}) as follows
	\begin{equation}
	\label{eq:Ps_nearest_1}
	\mathbb{P}_s^{\operatorname{nearest}} = \left(1+\xi^{-1} \tau^\delta\frac{\tilde{\lambda}_{\IoT}+\hat P_{\I}^\delta \tilde{\lambda}_{\I}}{\lambda_{\B}}\right)^{-1}.
	\end{equation}
	Thus, by setting $\mathbb{P}_s^{\operatorname{nearest}}\geq \epsilon$, we get the desired expression in (\ref{eq:UNB_lowerBound_Success}). We can also rewrite (\ref{eq:Ps}) for $\mathbb{P}_s\geq \epsilon$ in a similar manner.  $\hfill\blacksquare$
	
\begin{figure}[t!]
		\centering
		\begin{subfigure}[t]{.45\textwidth}
			\centering
			\includegraphics[width=\textwidth]{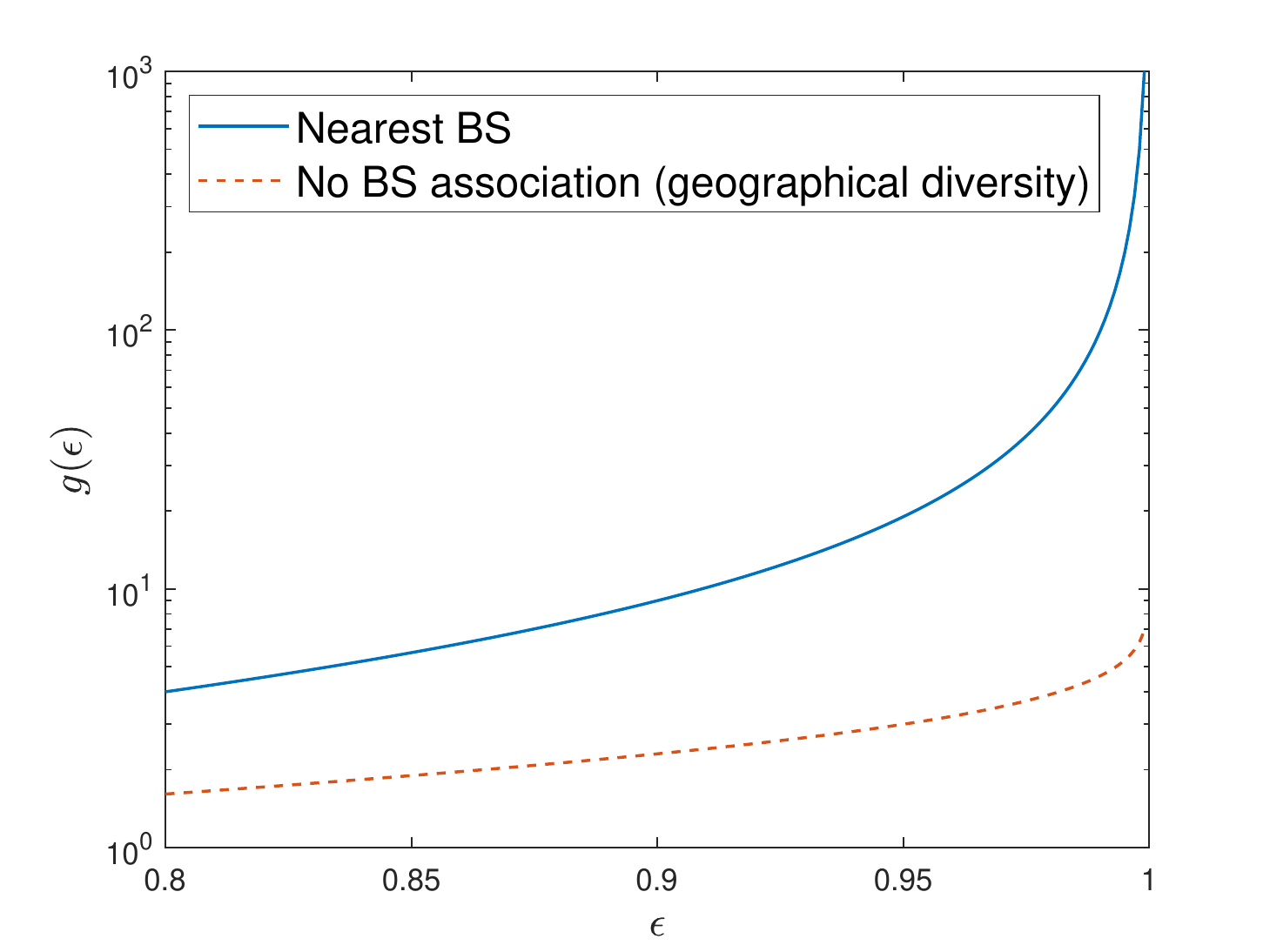}
			\caption{Lower bound for different $\epsilon$}
			\label{fig:UNB_lowerBound}
		\end{subfigure}
		\begin{subfigure}[t]{.45\textwidth}
			\centering
			\includegraphics[width=1\textwidth]{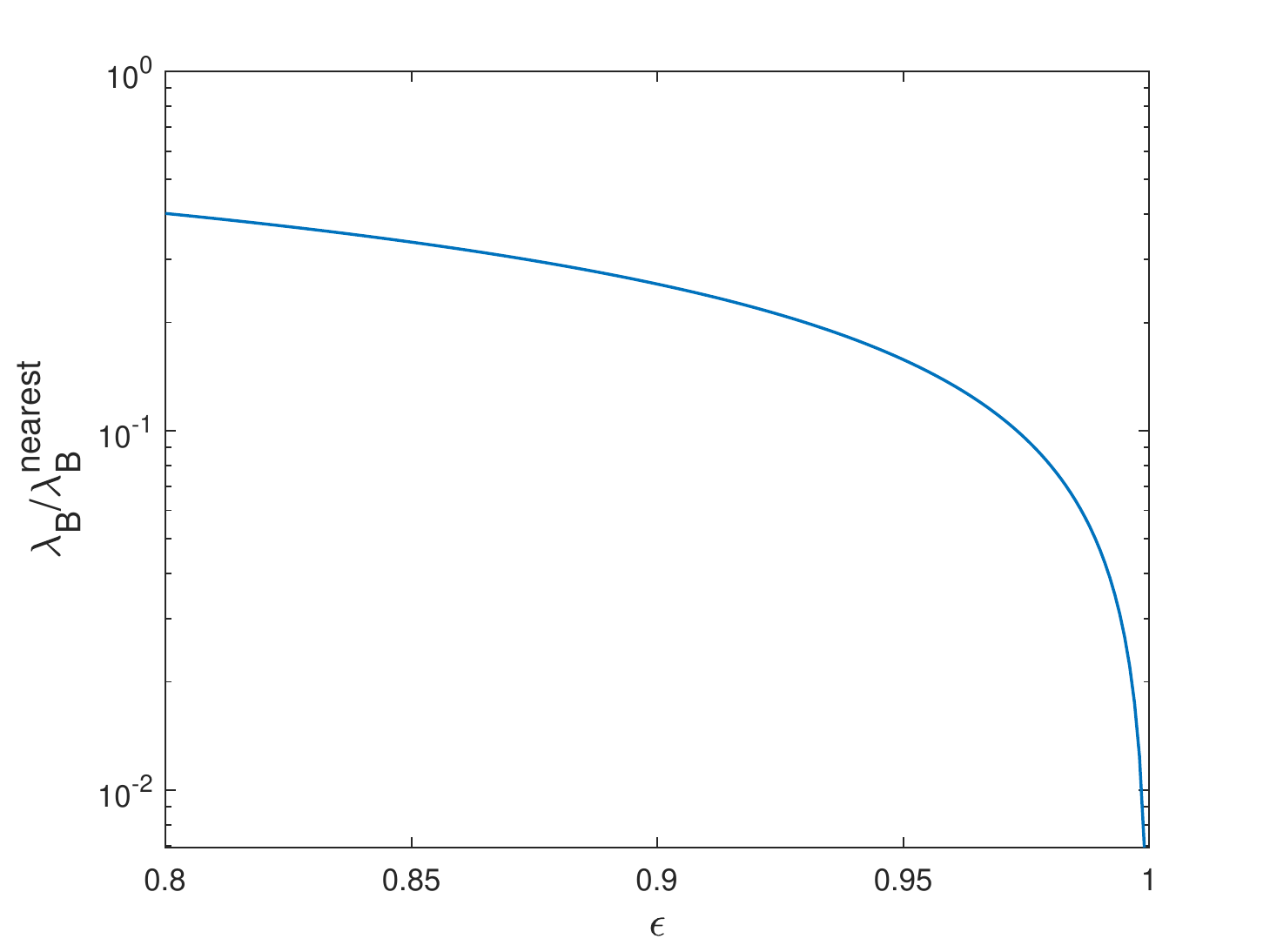}
			\caption{Density reduction for different $\epsilon$}
			\label{fig:UNB_BSratio}
		\end{subfigure}
		\caption{Impact of geographical diversity on the network resources needed, i.e., the number of bands and the density of BSs, to achieve a target success probability constraint.} 
		\label{fig:UNB_geographicalDiversity}
	\end{figure}

	We observe from Corollary \ref{corollary:geographicalDiversity} that as $\epsilon\rightarrow1$, then the lower bound in (\ref{eq:UNB_lowerBound_Success}) is significantly higher for nearest BS association, requiring a larger number of bands or UNB BSs. This is also shown in Fig. \ref{fig:UNB_lowerBound}. Put differently, if an IoT network operator relies on cell association and uses $\lambda_{\B}^{\operatorname{nearest}}$ BSs to achieve a target success probability of $\epsilon$, then switching to no BS association helps the operator reduce its BS deployment density to
	\begin{equation}
	\lambda_{\B} = \left(\frac{1-\epsilon}{\epsilon}\right) \ln\left(\frac{1}{1-\epsilon}\right) \lambda_{\B}^{\operatorname{nearest}}.
	\end{equation}
	In Fig. \ref{fig:UNB_BSratio}, we show the ratio $\lambda_{\B} /\lambda_{\B}^{\operatorname{nearest}}$ as $\epsilon\rightarrow1$. It is evident that geographical diversity reduces the density of BSs required by one to two orders of magnitude at high success probability constraints. 
	Likewise, we get similar trends by comparing the two schemes in terms of the transmission capacities in (\ref{eq:TC_nearest}) and (\ref{eq:TC}).
	
	%--------------------------------
	%Subsection: Optimizing the number of transmissions
	%--------------------------------
	\subsection{Optimizing the number of transmissions}
	
	In addition to the geographical diversity, UNB networks rely on repetition diversity, where the same packet is sent multiple times over time and/or frequency. For example, Sigfox typically uses three transmissions per packet; however, is this optimal in terms of maximizing the success probability? How many transmissions should each device send? We answer these questions by examining the impact of $N$ on the performance of the UNB network with no BS association.
	
	We can rewrite (\ref{eq:Ps}) as $\mathbb{P}_s =  1 - \exp\left(-\frac{ a_1 H_{N}}{a_2N+a_3}\right)$, where $a_1=\xi \tau^{-\delta} \lambda_{\B}$, $a_2=\beta_{T}\lambda_{\operatorname{T}} \cdot \frac{\beta_{F}b}{M\cdot B}\cdot \lambda_{\IoT}$, and $a_3=\hat P_{\I}^\delta\tilde{\lambda}_{\I}$. We can further show $H_N = \sum_{k=1}^N k^{-1}$ and since $P_s \propto H_N$, we can view $H_N$ as a gain in SINR. Remarkably, this gain is identical to the processing gain achieved under the selection combining scheme \cite{GhoshMuhamed2010}, i.e., repetition diversity is similar to selection combining as the BS does not coherently add the repeated transmissions. Nevertheless, sending the packet multiple times increases power consumption for a given payload (and latency, yet it is typically less critical), and it worsens the intra-network interference due to congesting the network with more transmissions. The following corollary presents the optimal $N$ that maximizes $\mathbb{P}_s$.
	
	\begin{corollary}\label{corollary:optimalN}
		The optimal number of transmissions that maximizes $\mathbb{P}_s$ is 
		\begin{equation}
		\label{eq:optimalN}
		N^\star = \operatorname*{argmin}_{N\in\mathbb{Z}} \left\{N \Big|(1+N) H_N - N > \frac{\hat P_{\I}^\delta\tilde{\lambda}_{\I}}{\beta_{T}\frac{\beta_{F}b}{M\cdot B}\lambda_{\operatorname{T}}\lambda_{\IoT}} \right\}.
		\end{equation}
		Further, in the absence of incumbents, then $N^\star =1$, i.e., the IoT device should send a single transmission per packet. 
	\end{corollary}
	\noindent \emph{Proof}: Let $f(N)=- \frac{ a_1 H_{N}}{a_2N+a_3}$. If $f(N+1)>f(N)$, then adding an additional transmission decreases the success probability. We can simplify the inequality as follows
	\begin{equation}
	\begin{aligned}
	- \frac{ a_1 H_{N+1}}{a_2(N+1)+a_3} &> - \frac{ a_1 H_{N}}{a_2N+a_3}\\
	%\frac{ a_1 H_{N+1}}{a_2(N+1)+a_3} &<  \frac{ a_1 H_{N}}{a_2N+a_3}\\
	\frac{H_{N+1}}{H_N} \cdot \frac{a_2N+a_3}{a_2N+a_3 + a_2}  &< 1. 
	\end{aligned}
	\end{equation}
	For the above inequality to hold, we must have
	\begin{equation}
	\frac{H_{N+1}}{H_N}  < 1+ \frac{1}{N+a_3/a_2}.
	\end{equation}
	Since $H_{N+1}=H_N + \frac{1}{N+1}$, we can simplify the above inequality to
	\begin{equation}
	\label{eq:N_condition}
	(1+N) H_N - N > \frac{a_3}{a_2}.
	\end{equation}
	Further, when the incumbents are absent, we have $a_3=0$, and thus the smallest $N$ that satisfies (\ref{eq:N_condition}) is $N=1$. $\hfill\blacksquare$
	
	It is observed from (\ref{eq:optimalN}) that the optimal number does not depend on the BS density or the decoding threshold. In addition, for Type-I incumbents, if $B_{\I,0} < M\cdot B$, then (\ref{eq:optimalN}) becomes independent of $M$, i.e., the optimal number of transmissions is independent of how many bands are used. This may not be the case for Type-II incumbents, as by definition of this model, adding more bands also adds more interfering networks.    
	
	Ideally, the number of transmissions should be reconfigurable, and a lower number of repetitions is recommended in regions where the UNB network is the dominant operator. Not only using lower $N$, if recommended, reduces power consumption of the device, but it can also widen the scope of services provided by the UNB network. Recall that Sigfox allows the IoT device to report a maximum of six unique packets per hour, as each one is sent three times.\footnote{In Europe, the UNB device has a duty cycle of 36s per hour; since each Sigfox transmission is 2s, the maximum number of transmissions is 18.} For applications that require constant monitoring with frequent status updates, using Sigfox becomes less appealing; however, if each device uses a single transmission, then the number of unique packets per hour increases to 18.
	
	%--------------------------------
	%Subsection: Random versus pseudorandom frequency hopping
	%--------------------------------
	\subsection{Random versus pseudorandom frequency hopping}\label{sec:PNhopping}
	
	Our analysis has considered the IoT device to randomly hop from one frequency to another. In this section, we consider a different system, where there is a predefined codebook of hopping patterns at a given time. These patterns are orthogonal, i.e., each pattern determines the $N$ channels to be used such that each one has channels orthogonal to all other patterns. The number of codes, over a given time slot, is equal to the number of channels; thus, the probability that two devices, transmitting with some time overlap, pick the same sequence is the same as the probability of two devices picking the same channel in the random hopping scheme.
	
	It can be shown that the densities of interfering IoT devices and incumbents remain the same even under the pseudorandom (PN) hopping scheme. However, in random hopping, each transmission could interfere with a different set of interfering IoT devices, whereas in PN hopping, if both devices pick the same hopping pattern, they will interfere in each of the $N$ transmissions. Thus, the question we want to answer in this section is the following: for a given density of interfering devices, do we want interfering devices to remain the same or to change across the $N$ transmissions? We answer this question in the following theorem. 
	
	\begin{theorem}\label{th:PNperformance}
		The success probability performance of the PN hopping scheme is upper bounded by that of the random hopping scheme for both nearest BS and no BS association systems.
	\end{theorem}
	\noindent \emph{Proof}: When the same devices interfere with each other in every transmission, then (\ref{eq:failProb}) becomes 
	\begin{equation}
	\label{eq:failProbPN}
	\begin{aligned}
	\mathbb{Q}_j^{\PN}  %&= \text{Pr}\left(\operatorname{SINR}_{1,j} \leq \tau, \operatorname{SINR}_{2,j} \leq \tau, \cdots, \operatorname{SINR}_{N,j}\leq \tau \right)\\
	&= \mathbb{E}_{\tilde \Phi_{\IoT},f_u} \left[\left(1-\mathbb{E}_{\tilde\Phi_{\I},f_k}\Big[e^{-\tau x_j^\alpha (\hat P_N+ I_{\operatorname{UNB}}+I_{\operatorname{INC}})}\Big]\right)^N\right]\\
	&\stackrel{(a)}{=}\mathbb{E}_{\tilde \Phi_{\IoT},f_u}\bigg[ \sum_{k=0}^N \binom{N}{k} (-1)^k e^{-k\pi \xi^{-1} (\tau\hat P_{\I})^\delta \tilde{\lambda}_{\I}x^2} \\ 
	&\times e^{-k\tau x_j^\alpha (\hat P_N + I_{\operatorname{UNB}})} \bigg]\\
	&\stackrel{(b)}{=}\sum_{k=0}^N \binom{N}{k} (-1)^k e^{-\pi \xi^{-1} \tau^\delta (k^\delta \tilde{\lambda}_{\IoT}+k\hat P_{\I}^\delta \tilde{\lambda}_{\I})x^2-k\tau x_j^\alpha \hat P_N} ,
	\end{aligned}
	\end{equation}
	where $(a)$ follows by taking the expectation with respect to $\tilde \Phi_{\I}$ and $f_k$ and then using the binomial theorem, and $(b)$ follows by taking the expectation with respect to $\tilde \Phi_{\IoT}$ and $f_u$. We note here that the set $\tilde \Phi_{\IoT}$ is the same across repetitions because UNB interference stems from the same set of UNB devices across repetitions. Following the same steps as in Appendix \ref{app:Ps}, the success probability in an interference-limited network and with nearest BS association is given as 
	\begin{equation}
	\begin{array}{ll}
	\label{eq:Ps_PN_nearest}
	\mathbb{P}_s^{\operatorname{nearest,PN}} =\\
	1-\textstyle \sum_{k=0}^N \binom{N}{k} (-1)^k \left(1+\xi^{-1} \tau^\delta\frac{k^\delta \tilde{\lambda}_{\IoT}+k\hat P_{\I}^\delta \tilde{\lambda}_{\I}}{\lambda_{\B}}\right)^{-1},
	\end{array}
	\end{equation}
	whereas the success probability with no BS association is 
	\begin{equation}
	\label{eq:Ps_PN}
	\mathbb{P}_s^{\operatorname{PN}} = 1 - \exp\left(\xi \tau^{-\delta}\sum_{k=1}^N \binom{N}{k} \frac{(-1)^k \lambda_{\B}}{k^\delta \tilde{\lambda}_{\IoT}+k\hat P_{\I}^\delta \tilde{\lambda}_{\I}}\right).
	\end{equation} 
	To prove that the performance of PN hopping is upper bounded by that of the random one, we can rewrite (\ref{eq:failProbPN}) as $\mathbb{E}[g(z)]$, where $g(z)=z^N$, and hence $(\ref{eq:failProbR})$ is rewritten as $g(\mathbb{E}[z])$. Using Jensen's inequality, we have $g(\mathbb{E}[z])\leq \mathbb{E}[g(z)]$, i.e., $\mathbb{Q}_j\leq \mathbb{Q}_j^{\PN}$, since $g(z)$ is a convex function. Thus, the success probability of the random hopping scheme outperforms that of the PN scheme for both nearest and no BS association. $\hfill\blacksquare$
	
	To summarize, random hopping inherently achieves an \emph{interference diversity} that can only improve the success probability performance. This diversity is attained because different transmissions experience interference from different sources. 
	
	\begin{figure*}[t!]
		\center
		\includegraphics[width=0.9\textwidth]{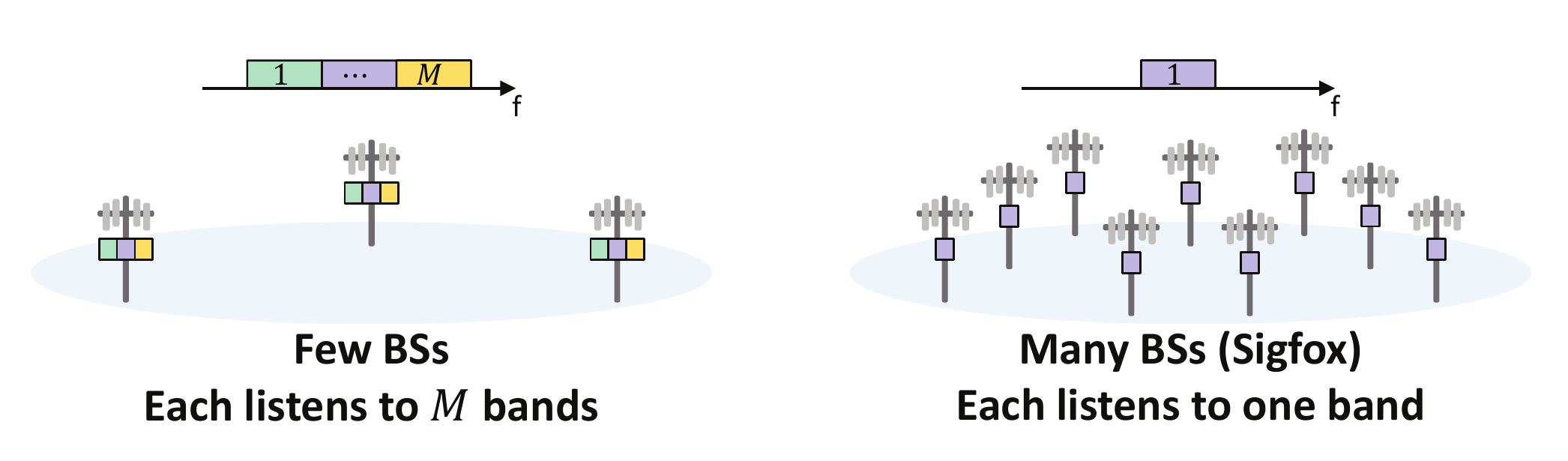}
		\caption{Two deployment approaches: few expensive BSs or many low-cost ones.}
		\label{fig:UNB_tradeoff}
	\end{figure*}

	%-----------------------------------------------------------
	%Section: Low-cost Multiband Access
	%-----------------------------------------------------------
	\section{Low-cost Multiband Access}\label{sec:UNB_optimization} 
	Corollary \ref{corollary:geographicalDiversity} has provided us with a lower bound on the UNB network resources needed to meet a success probability target of $\epsilon$. Interestingly, we observe that it is the product $M\lambda_{\B}$ that matters, i.e., we can trade the density of BSs with more bands. For example, the IoT network operator can halve the number of BSs, but then the network has to operate over twice the spectrum bandwidth. This observation demonstrates the interplay between frequency and space. In particular, by using a wider spectrum, the UNB device improves its chances to avoid interference from other UNB and incumbent devices. Alternatively, by deploying more BSs, the IoT device is likely to become closer to BSs in vicinity in addition to the increased geographical diversity.
	
	From an operator perspective, it is cheaper to deploy fewer BSs, while using a wider spectrum does not incur capital expenditure since the bands are unlicensed. However, the BS needs a more complex receiver if a wider spectrum is used. This follows because UNB networks are typically asynchronous in frequency, and thus the UNB receiver must sample the spectrum at a very high resolution to detect IoT signals. For this reason, existing UNB networks assume that devices send their signals within a single narrowband. Indeed, in the Sigfox network \cite{Sigfox2017}, the BS listens to a spectrum of bandwidth $B=200$KHz, where the power spectral density of the band is obtained using the Fast Fourier transform (FFT), with a very small sampling interval \cite{Artigue2017}. Specifically, consider a UNB device transmitting a signal of bandwidth $b=600$Hz in this multiplexing band and to detect the UNB signal, a resolution of $b/4$Hz is desired \cite{Artigue2017}. Then, the FFT size used by the BS must be at least $2^{12}$, which is on par with the FFT size used in Rel-15 5G New-Radio (NR) \cite{Jeon2018}. However, if the UNB BS is required to listen to a band of BW $M\cdot B=1$MHz, then the FFT size must be increased to $2^{14}$, increasing the FFT complexity by more than 4.5x.
	
	To summarize, if a UNB network operator wants to improve the success probability or increase the transmission capacity, it has two options, as shown in Fig. \ref{fig:UNB_tradeoff}. In the first one, the operator can keep the BS deployment density low, but then each BS requires very expensive receivers to scan $M$ bands. In the second one, the operator can use existing low-cost BSs, but it has to deploy them at a higher density. 
	
	One question we are interested to answer in this section is: can we improve the network performance if we use $M$ bands, but deploy low-cost BSs, where each one randomly selects one of the $M$ bands? To this end, we present two transmission protocols: Band-constrained and band-hopped multiband access. In the former, the IoT device sends all of its $N$ transmissions over one randomly selected band. In the latter, each transmission of the $N$ packets can be sent to a different band.   
	
\begin{figure*}[t!]
		\center
		\includegraphics[width=0.9\textwidth]{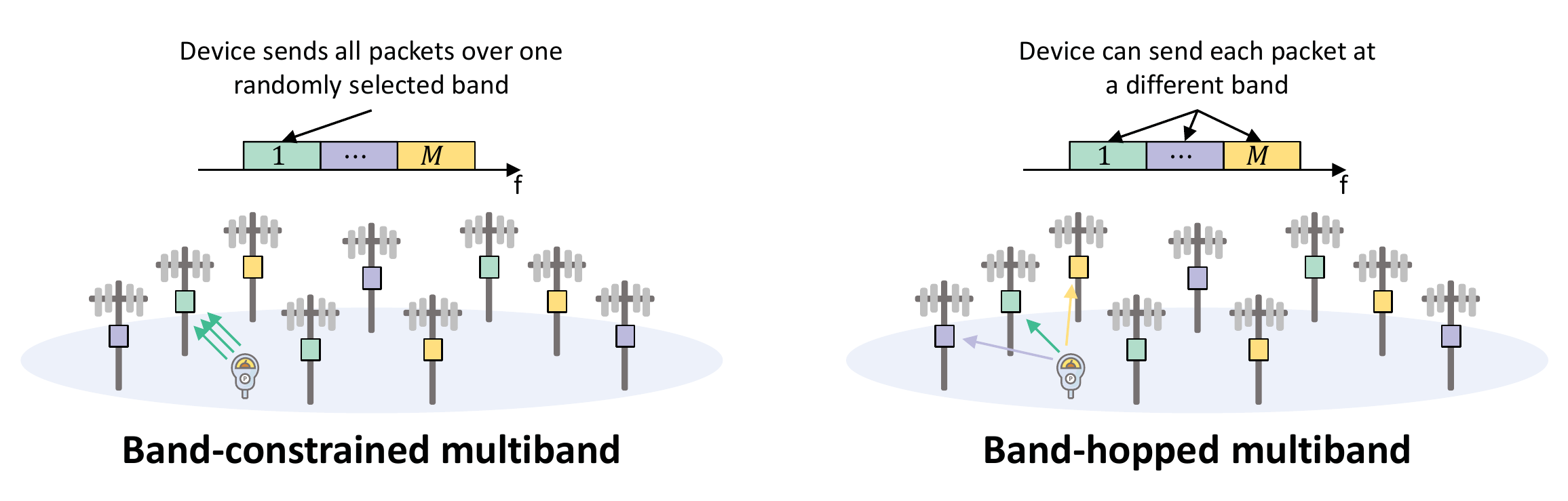}
		\caption{Two low-cost multiband access protocols: band-constrained and band-hopped.}
		\label{fig:UNB_multiband}
	\end{figure*}

	We note that the difference between the band-constrained and band-hopped multiband protocols is that in the former, the same set of BSs will listen to all $N$ messages, whereas in the latter, each message could be received at different subsets of BSs. These two low-cost multiband access protocols are illustrated in Fig. \ref{fig:UNB_multiband}. Next, we analyze the protocols under no BS association.
	
	%--------------------------------
	%Subsection: Band-constrained multiband with no BS association
	%--------------------------------
	\subsection{Band-constrained multiband with no BS association}\label{sec:SM}
	
	Let $\mathbb{P}_{s|m}$ denote the success probability given that the typical IoT device picks the $m$-th band for the transmission of $N$ messages. Then, the success probability is 
	\begin{equation}
	\label{eq:Ps_SM}
	\mathbb{P}_{s}^{\operatorname{BC}} = \frac{1}{M}\sum_{m=1}^M  \mathbb{P}_{s|m}.
	\end{equation} 
	Further, to generalize the analysis, let each BS picks the $M$-th band with probability $p_m$. Then, we have the following theorem. 
	\begin{theorem}\label{th:Ps_SM}
		In an interference-limited network, the success probability of the band-constrained multiband protocol with no BS association and Type-I incumbents is given as
		\begin{equation}
		\begin{array}{ll}
		\label{eq:SM_R_I}
		\mathbb{P}_s^{\operatorname{BC,I}} = \\
		1 -  \frac{1}{M}\sum_{m=1}^M  \exp\left(-\xi \tau^{-\delta} H_{N}\cdot \frac{p_m\lambda_{\B}}{\tilde{\lambda}_{\IoT}+\hat P_{\I}^\delta\min\{1,\frac{B_{\I,0}}{M\cdot B}\}\lambda_{\I,0}}\right),
		\end{array}
		\end{equation}
		whereas for Type-II incumbents, it is given as
		\begin{equation}
		\begin{array}{ll}
		\label{eq:SM_R_II}
		\mathbb{P}_s^{\operatorname{BC,II}} = \\
		1 -  \frac{1}{M}\sum_{m=1}^M  \exp\left(-\xi \tau^{-\delta} H_{N}\cdot \frac{p_m\lambda_{\B}}{\tilde{\lambda}_{\IoT}+\hat P_{\I}^\delta\frac{B_{\I,m}}{B}\lambda_{\I,m}}\right).
		\end{array}
		\end{equation}	
	\end{theorem}
	\noindent \emph{Proof}: We can compute $ \mathbb{P}_{s|m}$ as follows. The set of BSs listening to the $m$-th band is an HPPP process with density $p_m\lambda_{\B}$. Furthermore, the set of interfering IoT devices is the same as $\tilde \Phi_{\IoT}$. To show this, recall that the density of IoT devices with a time overlap is $N\beta_{T}\lambda_{\operatorname{T}}\lambda_{\IoT}$. Among these devices, only $1/M$ of them, on average, will select the $m$-th band. Among those that select the $m$-th band, only $\beta_{F} b/B$ will select, on average, the same channel as the typical IoT device. Finally, the set of interfering incumbents for Type-I remains the same as we assume the incumbent can use any part of the spectrum, whereas for Type-II, the set of interfering incumbents in the $m$-th band is $\frac{B_{\I,m}}{B}\lambda_{\I,m}$.  To this end, we can use the success probability of the benchmark protocol to compute $\mathbb{P}_{s|m}$, where we replace $\lambda_{\B}$ in (\ref{eq:Ps}) with $p_m \lambda_{\B}$ as well as replace the set of interfering incumbents with that corresponding to the interfering network's type. $\hfill\blacksquare$
	
	%--------------------------------
	%Subsubsection: Remarks about the protocol in presence of Type-I incumbents
	%--------------------------------
	\subsubsection{Remarks about the protocol in presence of Type-I incumbents}
	We note that $\mathbb{P}_s^{\operatorname{BC,I}}$ is maximized when $p_m=1/M$. This can be proved using the inequality $\frac{1}{M}\sum_{m=1}^M  \exp(p_m z)\geq \exp(\frac{z}{M} \sum_{m=1}^M p_m)=\exp(z/M)$, which holds with equality when $p_m=1/M$. This is intuitive as devices select a band with equal probability and incumbents can operate over any part of the spectrum; thus, UNB BSs must follow the same selection procedure to maximize the success probability performance. An interesting observation is in the absence of an incumbent network, this band-constrained multiband protocol has the same performance as the single-band protocol when $p_m=1/M$. This shows that while increasing the number of bands decreases the density of interfering devices, the density of BSs listening to the same band also decreases with the same rate, i.e., there is, on average, no gain of using more than one band in this case. However, in the presence of the incumbent network, increasing the number of bands can have a different impact, depending on the bandwidth of the interfering network. We can illustrate this in terms of the transmission capacity, which is given in the following corollary. 
	\begin{corollary}
		The transmission capacity of the band-constrained multiband protocol with Type-I incumbents and $p_m=1/M$ is 
		\begin{equation}
		\label{eq:TC_SM}
		\mathbb{C}^{\operatorname{BC,I}}(\gamma) = \textstyle\frac{\gamma  B}{\beta_{T}\beta_{F} b \lambda_{\operatorname{T}}}\left(\frac{\xi \tau^{-\delta} H_N\lambda_{\B}}{N\ln(\frac{1}{1-\gamma})}-\frac{\hat P_{\I}^\delta M\min\{1,\frac{B_{\I}}{M\cdot B}\}\lambda_{\I}}{N}\right).
		\end{equation}	
	\end{corollary}$\hfill\blacksquare$
	
	\noindent Let $\mathbb{C}^{\operatorname{SB}}(\gamma)$ denote the transmission capacity of the single-band, which is obtained by substituting $M=1$ in (\ref{eq:TC}). Further, assume Type-I incumbents, then we have the following remarks:
	\begin{itemize}
		\item If $B_{\I}<B$, then $\mathbb{C}^{\operatorname{BC,I}}(\gamma)=\mathbb{C}^{\operatorname{SB}}(\gamma)$. This follows because in both protocols there will be channels with no interfering incumbents. While the band-constrained multiband will have more of such channels, the density of BSs listening to a particular band is reduced.
		\item If $B_{\I}>M\cdot B$, then $\mathbb{C}^{\operatorname{BC,I}}(\gamma)<\mathbb{C}^{\operatorname{SB}}(\gamma)$. This follows because in both protocols the density of incumbents is the same and UNB devices cannot avoid them in frequency. However, in single-band access, all BSs use the same band, achieving higher geographical diversity compared to the band-constrained multiband protocol. Interestingly, this relation holds even when $B_{\I}>B$ and $B_{\I}<M\cdot B$, showing that for band-constrained multiband access, there is a cost due to randomly assigning each BS a single band to listen to, and thus increasing $M$ here can be harmful if the incumbent's bandwidth spans more than one band.
	\end{itemize}
	
	%--------------------------------
	%Subsubsection: Remarks about the protocol in presence of Type-II incumbents
	%--------------------------------
	\subsubsection{Remarks about the protocol in presence of Type-II incumbents}
	We can optimize band selection probabilities to maximize $\mathbb{P}_s^{\operatorname{BC,II}}$, which is achieved by solving the following problem
	\begin{equation}
	\label{eq:SMBopt}
	\begin{array}{cl}
	\underset{\{p_m\}}{\text{minimize}} 
	&~~ \sum_{m=1}^M \exp(-c_m p_m),\\
	\text{subject to}     
	&~~\sum_{m=1}^M p_m=1, ~~~p_m\geq0,~~m=1,2,\ldots M,\\
	\end{array}
	\end{equation}
	where $c_m = \frac{\xi \tau^{-\delta} H_{N}\lambda_{\B}}{\tilde{\lambda}_{\IoT}+\hat P_{\I}^\delta\frac{B_{\I,m}}{B}\lambda_{\I,m}}$. This is a convex problem and its optimal solution is
	\begin{equation}
	\label{eq:optPSM}
	p_m^\star = \max\left\{0,\frac{1}{c_m}\ln\left(\frac{c_m}{\nu^\star}\right)\right\},
	\end{equation}
	where $\nu^\star$ is the Lagrange multiplier for the equality constraint. In Appendix \ref{app:dual}, we prove that  $\nu^\star$ satisfies 
	\begin{equation}
	\sum_{m=1}^{M} \max\left\{0,\frac{1}{c_m}\ln\left(\frac{c_m}{\nu^\star}\right)\right\}  = 1. 
	\end{equation}
	
	We have the following remarks about the solution in (\ref{eq:optPSM}). First, if the incumbents are homogeneous across bands, i.e., $c_m = \bar c~ \forall m$, then, as expected, $p_m^\star = 1/M$. Second, the selection probabilities $\{p_m\}$ are not directly proportional to incumbents' densities. Indeed, the relationship is $p_m \propto \frac{1}{c_m}\log(c_m/\nu)$, i.e., $p_m$ increases as $c_m \rightarrow e^1\cdot \nu$ and then decreases beyond that point. In other words, when a band does not have many interfering devices, then $p_m$ could be high for that band. If then the number of interfering devices increased in that band, then more UNB BSs should switch to that band to compensate for the increased interference, up to a certain point for which beyond it, the UNB network should start reducing the presence of their BSs in that band. Finally, the selection probabilities require prior knowledge about the densities of incumbent networks and their BWs' occupancy. Such knowledge could be obtained via databases \cite{Khan2017}, spectrum measurements \cite{Ding2015}, or using machine learning \cite{Lees2019}. If it is infeasible to acquire knowledge about incumbents, then it becomes more reasonable to let the BS randomly select a band similar to the band selection probabilities of IoT devices. 
	
	%--------------------------------
	%Subsection: Band-hopped multiband with no BS association
	%--------------------------------
	\subsection{Band-hopped multiband with no BS association} 
	
	In this protocol, each packet can be sent over a different multiplexing band. Assuming that the typical device sends $n_m$ of the $N$ packets over the $m$-th band, where $\sum_{m=1}^M n_m= N$, then the probability of having at least a single message decoded successfully at any BS with Type-I incumbents is given as
	\begin{equation}
	\begin{array}{ll}
	\mathbb{P}_{s|\{n_m\}_{1}^M}^{\operatorname{I}} = \\
	1 - \prod_{m=1}^M \exp\left(-\xi \tau^{-\delta} H_{n_m}\cdot  \frac{p_m\lambda_{\B}}{\tilde{\lambda}_{\IoT}+\hat P_{\I}^\delta\min\{1,\frac{B_{\I,0}}{M\cdot B}\}\lambda_{\I,0}}\right),
	\end{array}
	\end{equation}
	and for Type-II incumbents, it is given as 
	\begin{equation}
	\begin{array}{ll}
	\mathbb{P}_{s|\{n_m\}_{1}^M}^{\operatorname{II}} = \\
	1 - \prod_{m=1}^M \exp\left(-\xi \tau^{-\delta} H_{n_m}\cdot  \frac{p_m\lambda_{\B}}{\tilde{\lambda}_{\IoT}+\hat P_{\I}^\delta\frac{B_{\I,m}}{B}\lambda_{\I,m}}\right).
	\end{array}
	\end{equation}
	These expressions follow because the $m$-th band has $p_m \lambda_{\B}$ BSs listening to $n_m$ messages. Let $\mathcal{N}=\left\{n_1,n_2,\cdots,n_M|\sum_{m=1}^M n_m=N, n_m\in\{0,1,\cdots,N\}\right\}$ be the set of all possible combinations of sending $N$ messages over $M$ bands. Then, the success probability of this protocol is computed by averaging over $\mathcal{N}$, as given in the following theorem. 
	
	\begin{theorem}\label{th:Ps_UM}
		In an interference-limited network, the success probability of the band-hopped multiband protocol with no BS association is given as
		\begin{equation}
		\label{eq:UM_R}
		\mathbb{P}_s^{\operatorname{BH},\mathcal{I}} = \frac{1}{M^N}\sum_{\{n_m\}\subset\mathcal{N}} \frac{N!}{n_1!n_2!\cdots  n_M!}\mathbb{P}_{s|\{n_m\}_{1}^M}^{\mathcal{I}},
		\end{equation}
		where $\mathcal{I}\in\{\operatorname{I,II}\}$ and the sum is over every possible combination in $\mathcal{N}$.
	\end{theorem}
	\noindent \emph{Proof}: The probability of sending $n_m$ packets out of $N$ over the $m$th band follows the multinomial distribution, where $N$ is the number of trials and $M$ is the number of possible outcomes for each one. Since each channel is picked with probability $1/M$, then a specific combination $\{n_m\}_{m=1}^M\subset\mathcal{N}$ occurs with probability $\frac{1}{M^N} \frac{N!}{n_1!n_2!\cdots  n_M!}$, which completes the proof. $\hfill\blacksquare$
	
	We can optimize $p_m$ when the interfering incumbents are heterogeneous across bands. Similar to the formulation in (\ref{eq:SMBopt}), we can show that maximizing $\mathbb{P}_s^{\operatorname{BH,II}}$ is equivalent to solving the following convex problem 
	\begin{equation}
	\label{eq:UMBopt}
	\begin{array}{cl}
	\underset{\{p_m\}}{\text{minimize}} 
	&~~ \sum_{i} \exp\left(-\sum_{m=1}^M a_{m,i} p_m + d_i\right),\\
	\end{array}
	\end{equation}
	subject to the same constraints in (\ref{eq:SMBopt}). Here, the outer sum is with respect to every possible combination in $\mathcal{N}$, where $a_{m,i} = \frac{\xi \tau^{-\delta} H_{{i_{m}}}\lambda_{\B}}{\tilde{\lambda}_{\IoT}+\hat P_{\I}^\delta\frac{B_{\I,m}}{B}\lambda_{\I,m}}$ and $d_i = \frac{N!}{i_1!i_2!\cdots  i_M!}$. We solve this problem using a standard convex optimization solver \cite{GrantBoyd2016}.

	%-----------------------------------------------------------
	%Section: Simulation Results
	%-----------------------------------------------------------
	\section{Simulation Results}\label{sec:UNB_simulations}
	
	We run Monte Carlo simulations to validate the theoretical expressions. Specifically, we first generate BSs, UNB devices, and interfering devices according to their respective HPPPs. Then, we generate traffic for each device, determining the time and frequency stamps of each packet. For each link between a device and a BS, we generate fading channels and then evaluate the SINR of each packet either at the nearest BS or at all BSs to determine if the packet is successfully decoded. We repeat this process for  $10^4$ spatial realizations to get the success probability for each protocol.
	
	Unless otherwise stated, we use the simulation parameters given in Table \ref{tab:UNB_parameters}, where the UNB network emulates the Sigfox network with US specifications. The temporal traffic generation assumes that each device sends six unique packets per hour, each with a duration of $T\approx 0.35$s. For interfering incumbents, we consider LoRa IoT devices \cite{Bor2016} and assume they have a similar temporal traffic generation. Unless otherwise states, Type-II incumbents are homogeneous across bands, with both Type-I and Type-II incumbents being identical in single-band access. We note that the thermal noise is not ignored, i.e., $P_N\neq 0$ in Monte Carlo simulations. Finally, we use markers to denote Monte Carlo simulations and lines to denote the theoretical results. 
	
	\begin{table}[!t]
		\caption{Main parameters}
		\label{tab:UNB_parameters}
		\centering
		\scriptsize
		\begin{tabular}{|l|l|}
			\hline
			\textbf{Description }  				&  \textbf{Parameters}\\\hline
			IoT signal bandwidth				&$b=600$Hz	\\	
			UNB multiplexing band   			&$B=200$KHz\\
			Number of unique packets per hour 	& $K=6$ packets\\
			Packet size							&26 bytes\\
			Transmission duration   			&$T=26\times 8/b$\\
			Number of transmissions 			& $N=3$\\
			Number of bands         			& $M=5$\\
			IoT Tx power 						& $P_{\IoT}=14$dBm\\
			IoT density						    & $\lambda_{\IoT}=30\times 10^3/\lambda_{\B}$\\
			Incumbent bandwidth 				& $B_{\I,0}=B_{\I,m}=125$KHz\\
			Incumbent effective density 		& $\lambda_{\I,0}= \lambda_{\I,m} = 10^3 \lambda_{\operatorname{T}}/\lambda_{\B}$\\
			Incumbent Tx power 					& $P_{\I}=14$dBm (over $B_{\I,0}$ or $B_{\I,m}$)\\
			Noise power 						& $P_N=-146$dBm (over $b$)\\
			Path loss exponent 					&$\alpha=3.5$\\\hline
		\end{tabular}
	\end{table}

	%--------------------------------
	%Subsection: Impact of synchronization and hopping patterns
	%--------------------------------
	\subsection{Impact of synchronization and hopping patterns}
	
\begin{figure}[t!]
		\center
		\includegraphics[width=0.5\textwidth]{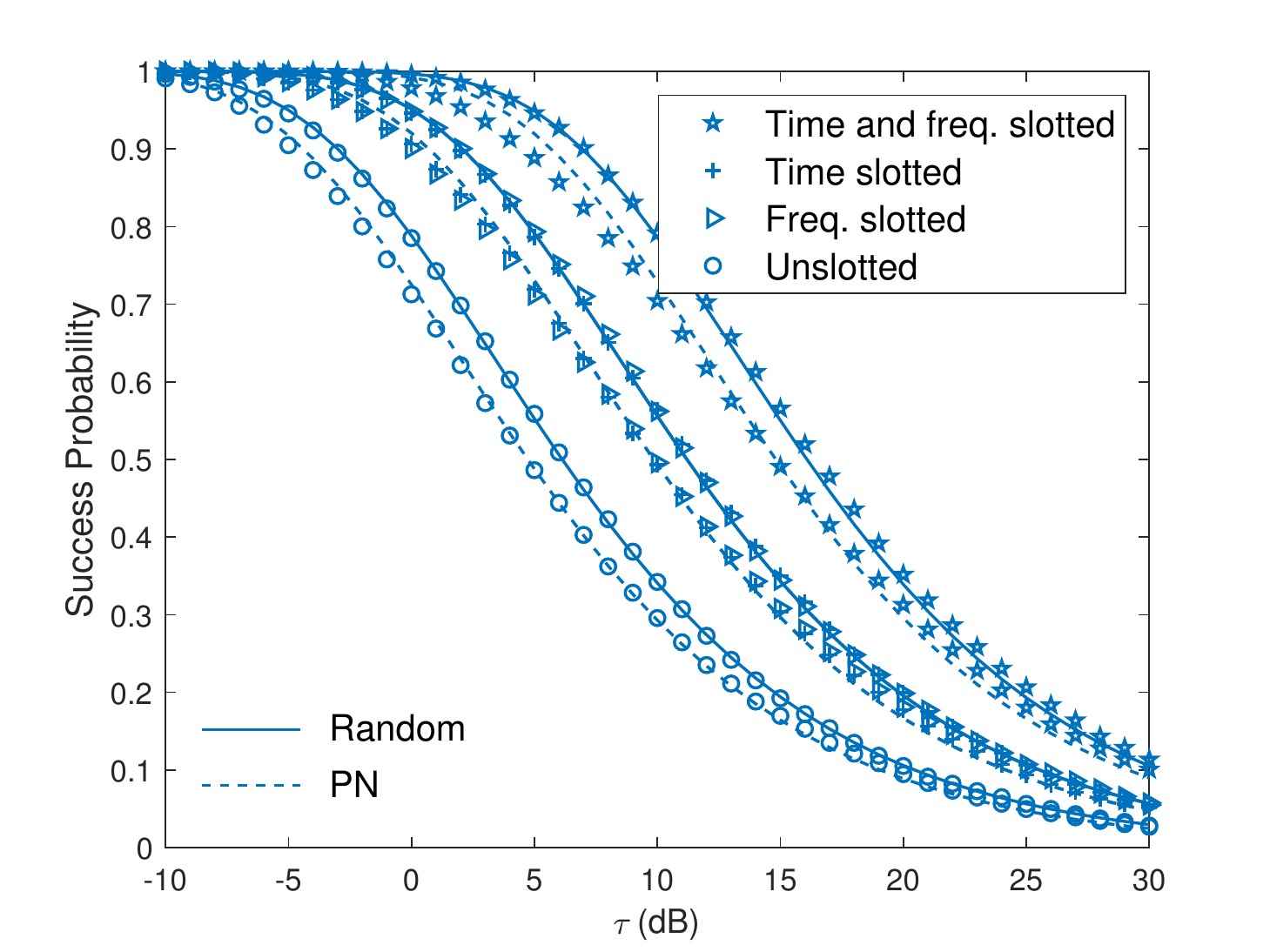}
		\caption{Success probability for Sigfox under different access cases ($M=1$ and $N=3$)}
		\label{fig:Ps_vs_tau_accessCases}
	\end{figure}

	We first consider the single-band Sigfox protocol and study the success probability performance under different hopping patterns and access cases, i.e, different levels of time-frequency synchronization. Fig. \ref{fig:Ps_vs_tau_accessCases} shows the success probability with variations of the SINR threshold. We have the following observations. First, the theoretical analysis matches well with simulations. Second, a time-frequency synchronized protocol tangibly improves the performance over a completely asynchronous protocol, e.g., the median SINR approximately improves by 10dB. Third, a time-slotted protocol has identical performance to a frequency-slotted one. Since UNB networks rely on narrowband signals, it is easier to achieve synchronization in time, and it is worth exploring due to the achieved SINR improvements over the unslotted system, e.g., the median SINR increases by approximately 5dB. Last, random repetition outperforms PN repetition by approximately 1dB because in the former there is \emph{an interference diversity gain} achieved as each transmission can experience a different set of interferers compared to the PN scheme. In the next results, we only focus on the unslotted time-frequency system, with random hopping patterns.
	
	%--------------------------------
	%Subsection: Success probability comparison
	%--------------------------------
	\subsection{Success probability comparison}
	
\begin{figure}[t!]
		\centering
		\begin{subfigure}[t]{.45\textwidth}
			\centering
			\includegraphics[width=\textwidth]{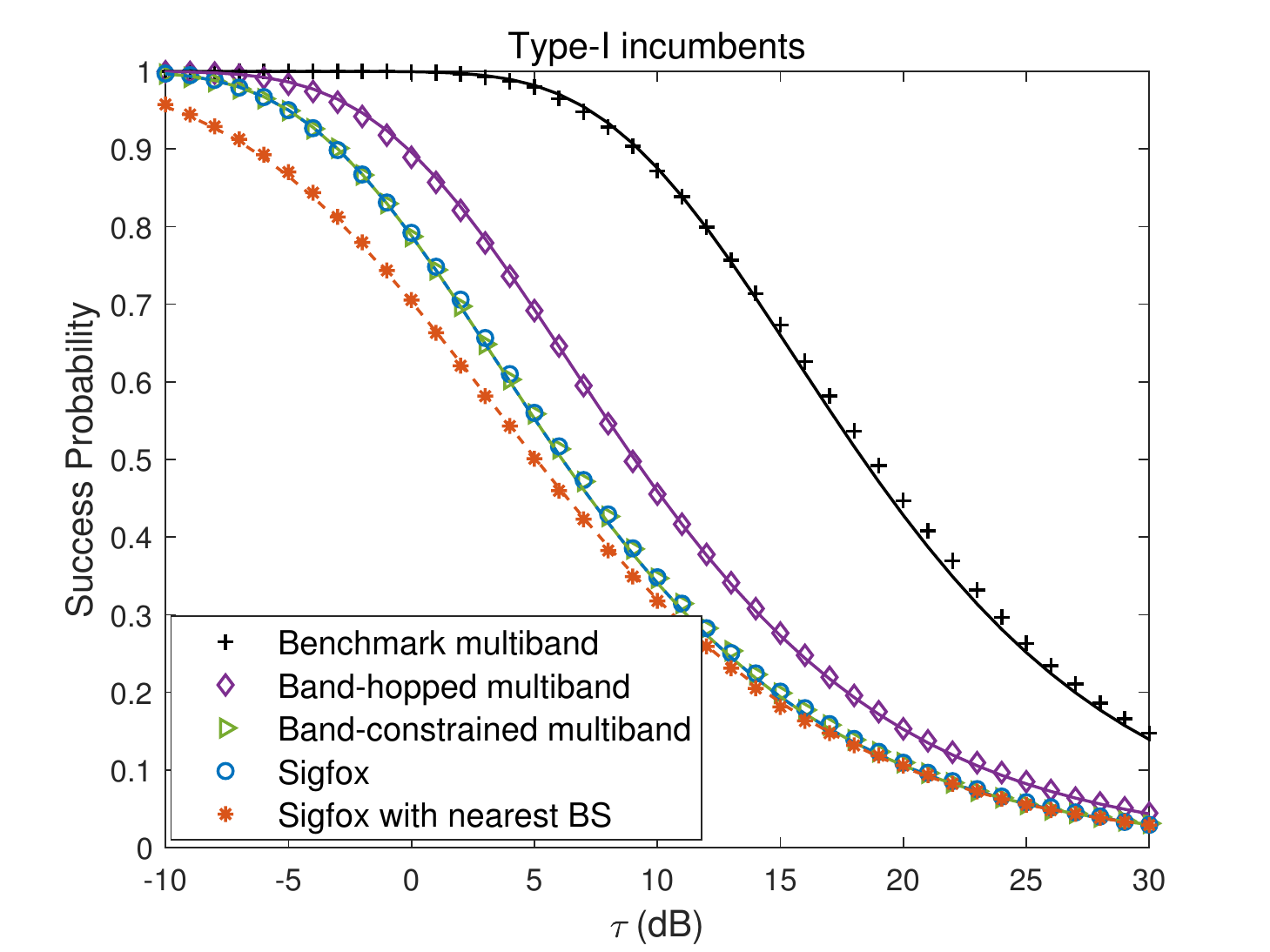}
			\caption{Type-I incumbents}
			\label{fig:Ps_comparison_I}
		\end{subfigure}
		\begin{subfigure}[t]{.45\textwidth}
			\centering
			\includegraphics[width=1\textwidth]{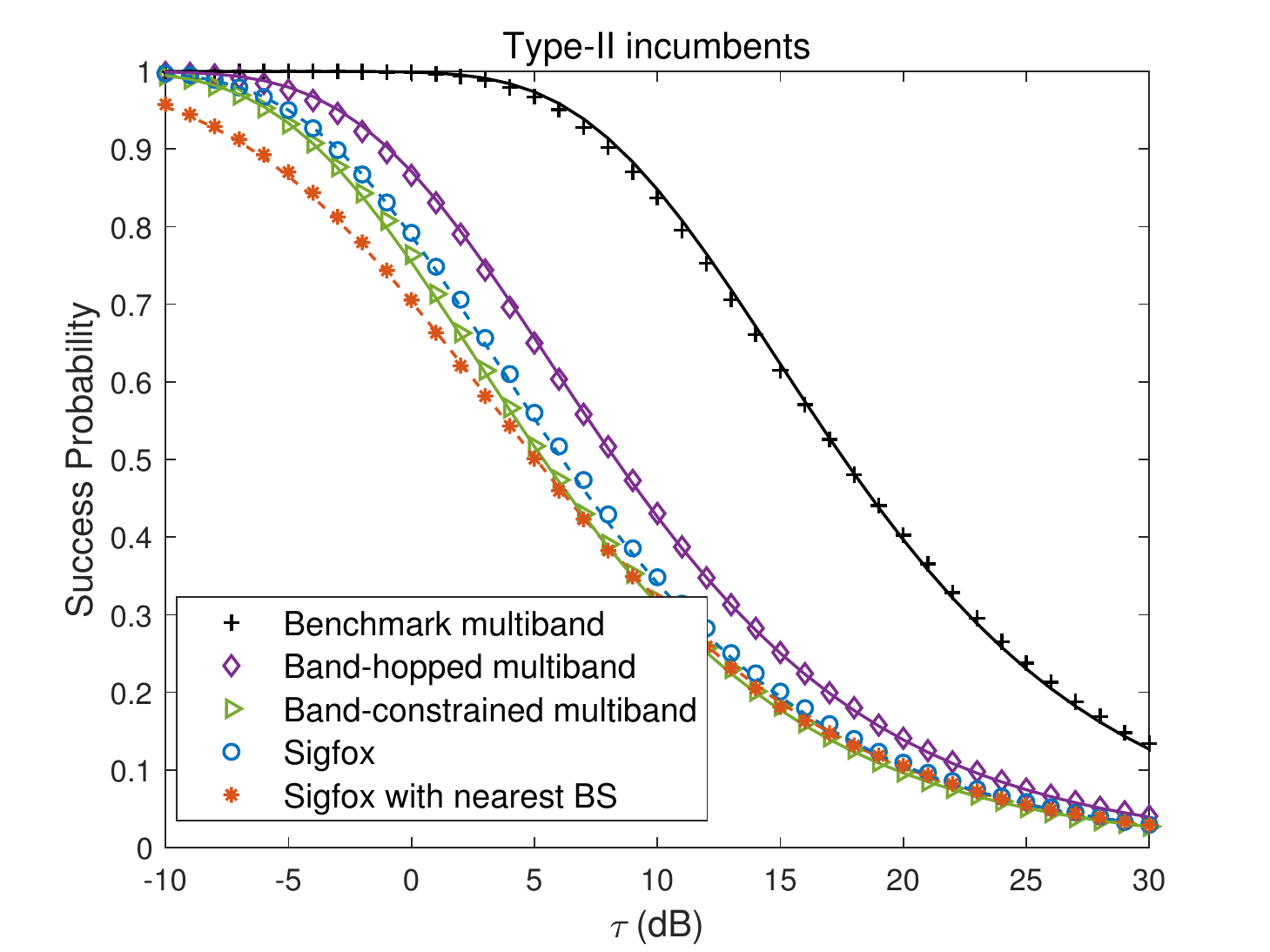}
			\caption{Type-II incumbents (homogeneous)}
			\label{fig:Ps_comparison_II}
		\end{subfigure}
		\caption{Success probability comparison of different UNB protocols ($M=5$ and $N=3$).}
		\label{fig:Ps_comparison}
	\end{figure}

	We compare the performance of several UNB protocols in terms of the success probability with variations of the SINR threshold. We consider the following schemes:
	\begin{itemize}
		\item  \textbf{Benchmark multiband}: Each BS listens to all $M$ bands, requiring complex receivers.
		\item \textbf{Band-hopped multiband}: Unless otherwise stated, a BS randomly selects a band and the IoT device sends its packets, each possibly over a different band. We also refer to this protocol as the proposed multiband protocol.   
		\item \textbf{Band-constrained multiband}: This is similar to the band-hopped protocol, except that the IoT device sends all its packets over a single randomly selected band. 
		\item \textbf{Sigfox}: All BSs and IoT devices use the same single band.
		\item \textbf{Nearest BS}: This is similar to Sigfox, but the IoT device only communicates with the nearest BS. The control overhead for cell association is ignored.
	\end{itemize}
	The success probability under Type-I incumbents is shown in Fig. \ref{fig:Ps_comparison_I}. We have the following remarks. First, the benchmark scheme significantly improves the success probability compared to Sigfox, e.g., the median SINR improves by approximately 12dB. However, the benchmark is impractical to implement due to the high computational complexity of processing a wider band at a very fine resolution. The proposed band-hopped protocol provides a practical compromise, where the median SINR improves by 3dB relative to Sigfox, at virtually no hardware cost. Further, the geographical diversity achieved with no BS association is particularly beneficial for cell-edge users, e.g., the cell-edge SINR in Sigfox with no BS association is 5dB higher than that of Sigfox with nearest BS association, and it is further improved by an additional 3dB when the proposed band-hopped protocol is used. Finally, the band-constrained protocol is worse than the band-hopped one, and in fact the former has identical performance to Sigfox, as explained in Section \ref{sec:SM}. This follows because in the band-hopped protocol, each packet can be received by different sets of BSs, whereas in the band-constrained one, all packets are received by the same set of BSs. Thus, the former protocol provides an additional geographical diversity gain over the latter, although each band has the same density of BSs under these two protocols.
	
	The success probability under Type-II incumbents is shown in Fig. \ref{fig:Ps_comparison_II}. The performance of single-band protocols remains the same, as we assume homogeneous interfering networks. For the benchmark and the proposed protocols, we still observe tangible performance gains over Sigfox, albeit the gains are slightly lower than those under Type-I incumbents because all bands have interfering incumbents under Type-II. Interestingly, band-constrained multiband performs slightly worse than Sigfox because each band is occupied by incumbents, while the density of BSs in each band is lower compared to Sigfox. This is not an issue in the band-hopped multiband due to the additional geographical diversity achieved by sending packets across different bands, which can be received by different sets of BSs. 
	
\begin{figure}[t!]
		\centering
		\begin{subfigure}[t]{.45\textwidth}
			\centering
			\includegraphics[width=\textwidth]{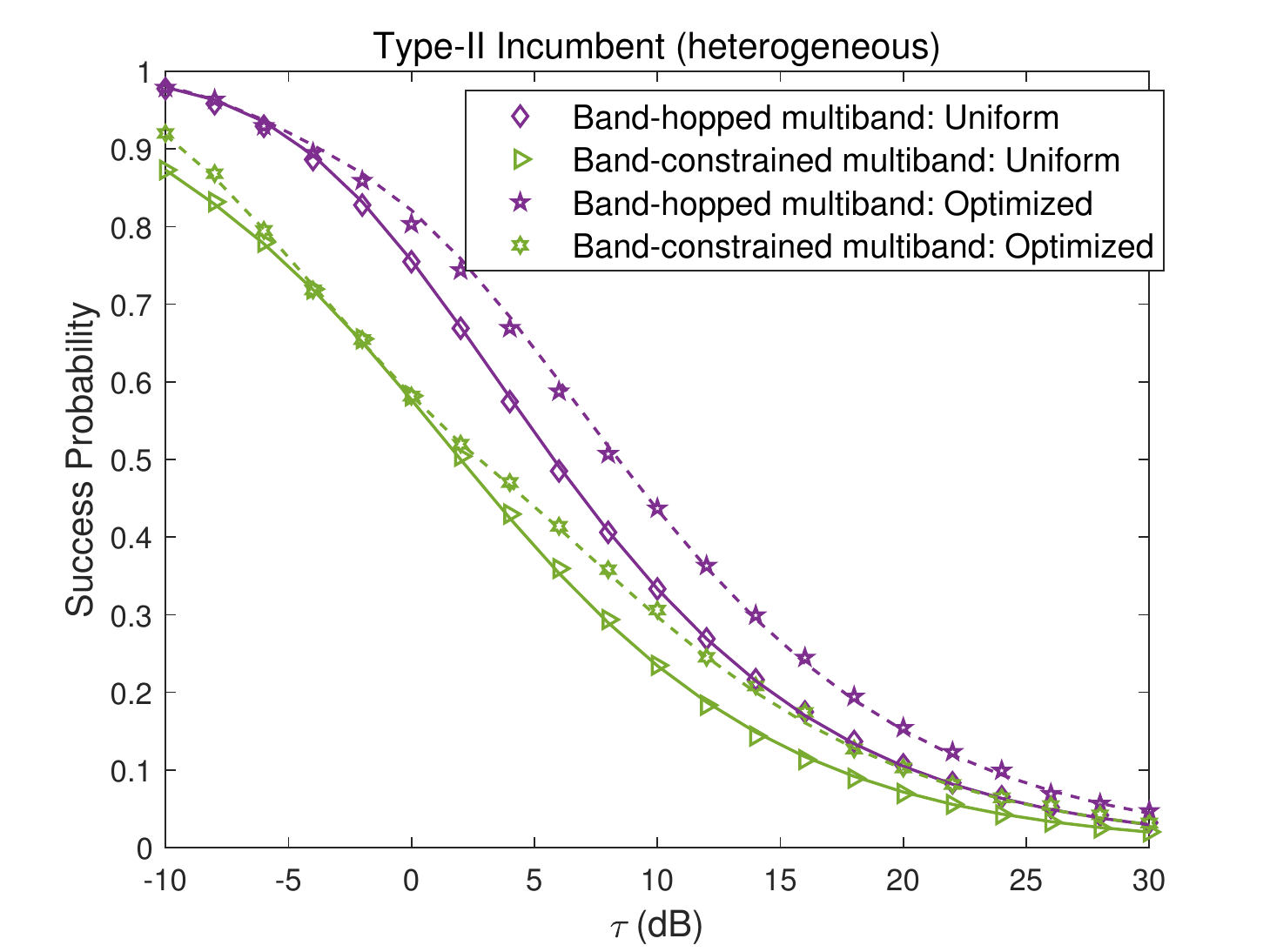}
			\caption{Success probability performance}
			\label{fig:Ps_comparison_II_het}
		\end{subfigure}
		\begin{subfigure}[t]{.45\textwidth}
			\centering
			\includegraphics[width=1\textwidth]{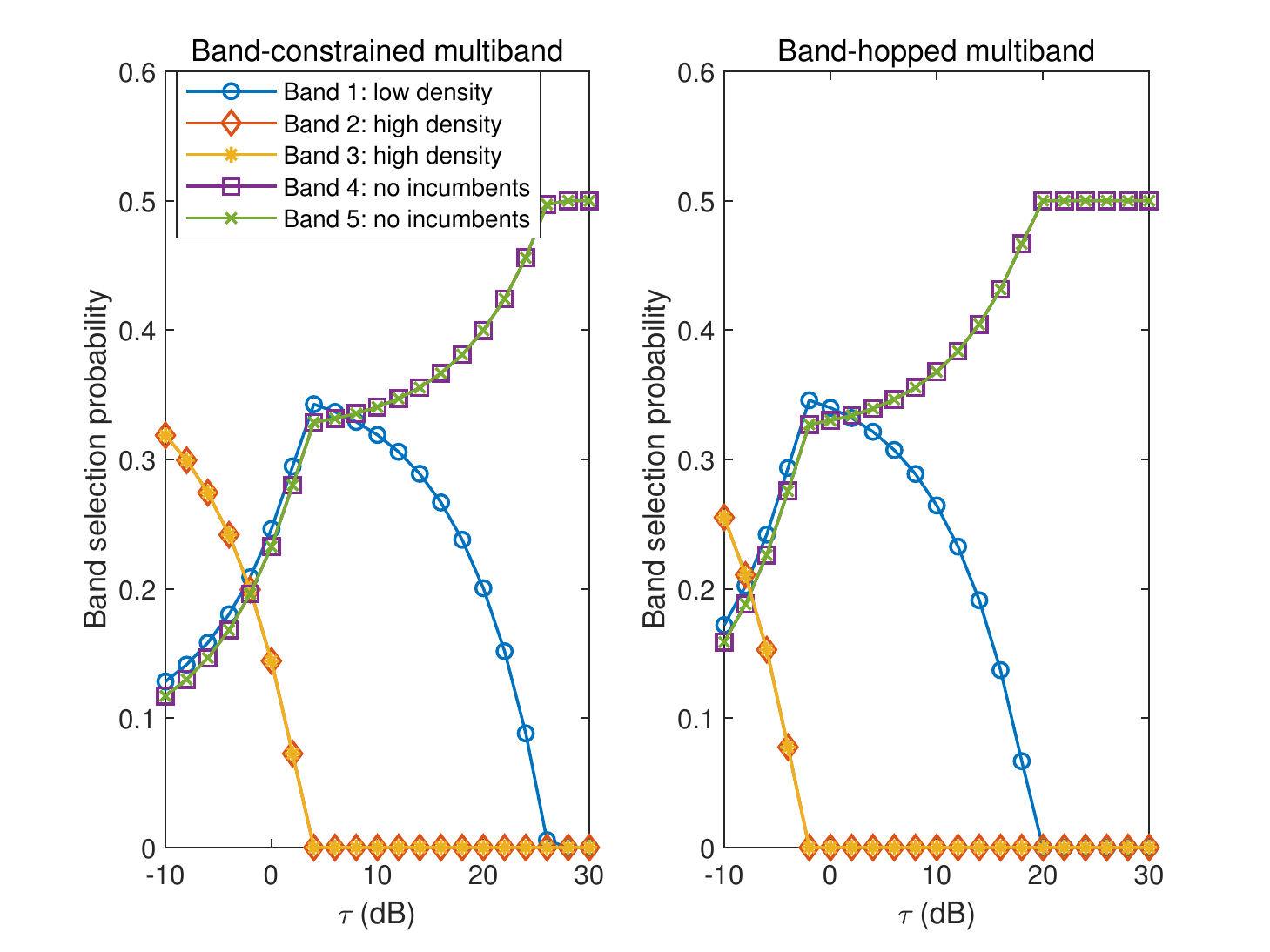}
			\caption{Optimal band selection probability}
			\label{fig:pm_vs_tau}
		\end{subfigure}
		\caption{Performance of multiband protocols with and without optimizing band selection.}
		\label{fig:Ps_comparison_het}
	\end{figure}

	We then study the performance of the multiband protocols in the presence of heterogeneous incumbents, where $M=5$ and $N=3$. Here, we assume the first band has 1000 LoRa devices/BS as before, the second and third bands have 30,000 devices/BS, and the last two bands do not have any incumbent devices. We study the band-constrained and band-hopped multiband protocols with uniform band selection, i.e., $p_m = 1/M$, and optimized band selection based on the solutions of (\ref{eq:SMBopt}) and (\ref{eq:UMBopt}). Fig. \ref{fig:Ps_comparison_II_het} shows the success probability performance. It is observed that optimizing the band selection probability can provide SINR gains when the interfering networks differ across bands. For instance, the median SINR improves by approximately 2dB for the band-hopped multiband. For the band-constrained scheme, the gains are primarily observed at high SINR. The optimal selection probability for each band is shown in Fig. \ref{fig:pm_vs_tau}. It is observed that when the decoding threshold, $\tau$, is low, then more BSs select the bands with higher inter-network interference to compensate for the additional interference. For bands with lower density of incumbents, the interference is not high and since the decoding threshold is low, then fewer BSs are sufficient to provide coverage. However, as the decoding threshold increases, higher SINR is needed for successful transmission; thus, the high interference in the busier bands may make it infeasible to reach that threshold without adding more BSs and affecting the performance in other bands. For this reason, we observe that for high $\tau$, UNB BSs start switching to the bands with lower density of interfering incumbents. Finally, comparing the band-hopped protocol to the band-constrained one, we observe that the UNB BS, under the former, is more likely to select a band with a lower incumbent density because the IoT device transmits its packets over multiple bands instead of a single one, i.e., there is a higher chance the IoT device transmits over a lightly occupied band compared to the band-constrained multiband. 
	
	%--------------------------------
	%Subsection: Impact of the number of bands and number of repetitions
	%--------------------------------
	\subsection{Impact of the number of bands and number of repetitions}
	
\begin{figure}[t!]
		\centering
		\begin{subfigure}[t]{.45\textwidth}
			\centering
			\includegraphics[width=\textwidth]{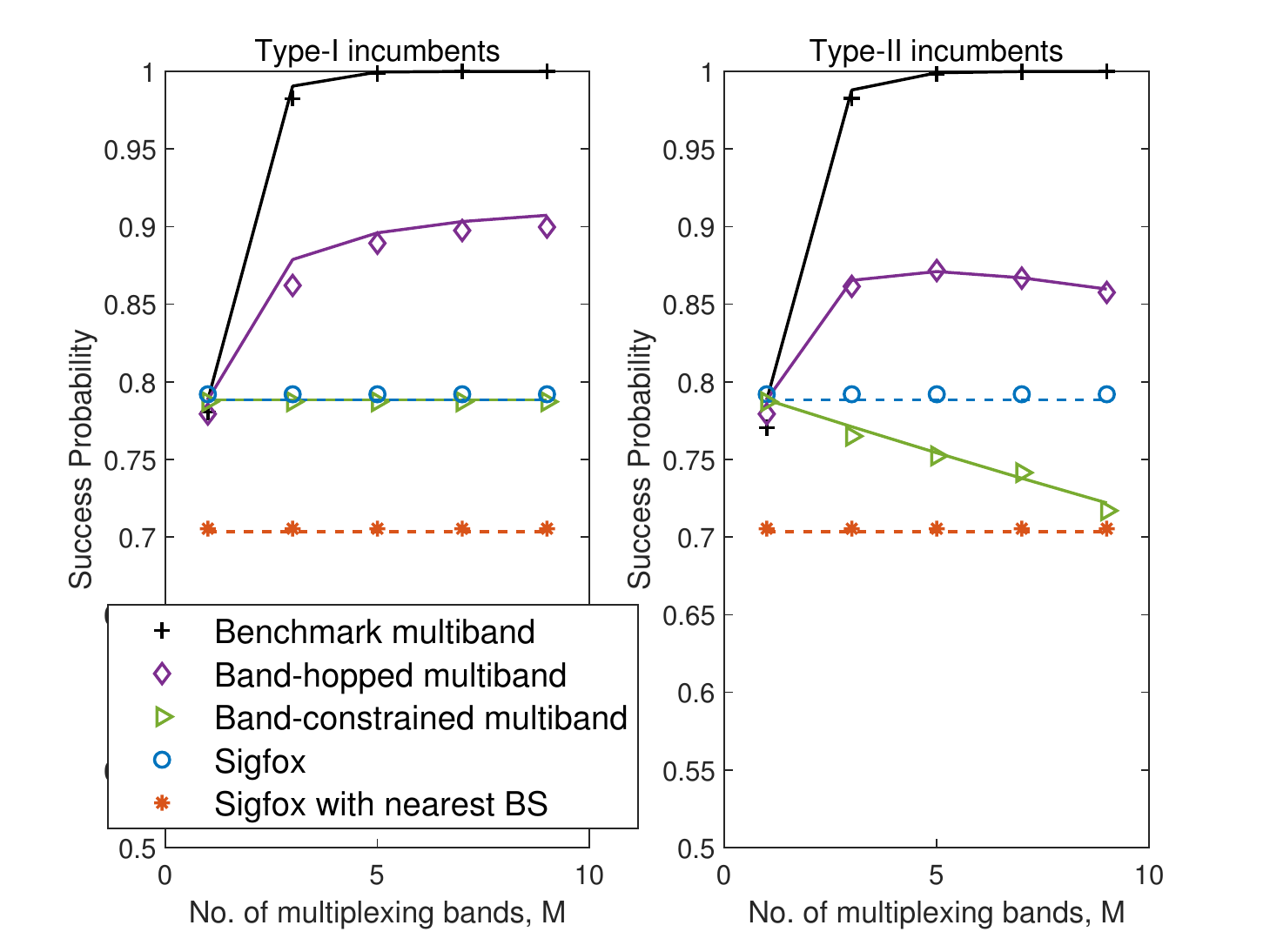}
			\caption{Variations of $M$ ($N = 3$)}
			\label{fig:Ps_vs_M}
		\end{subfigure}
		\begin{subfigure}[t]{.45\textwidth}
			\centering
			\includegraphics[width=1\textwidth]{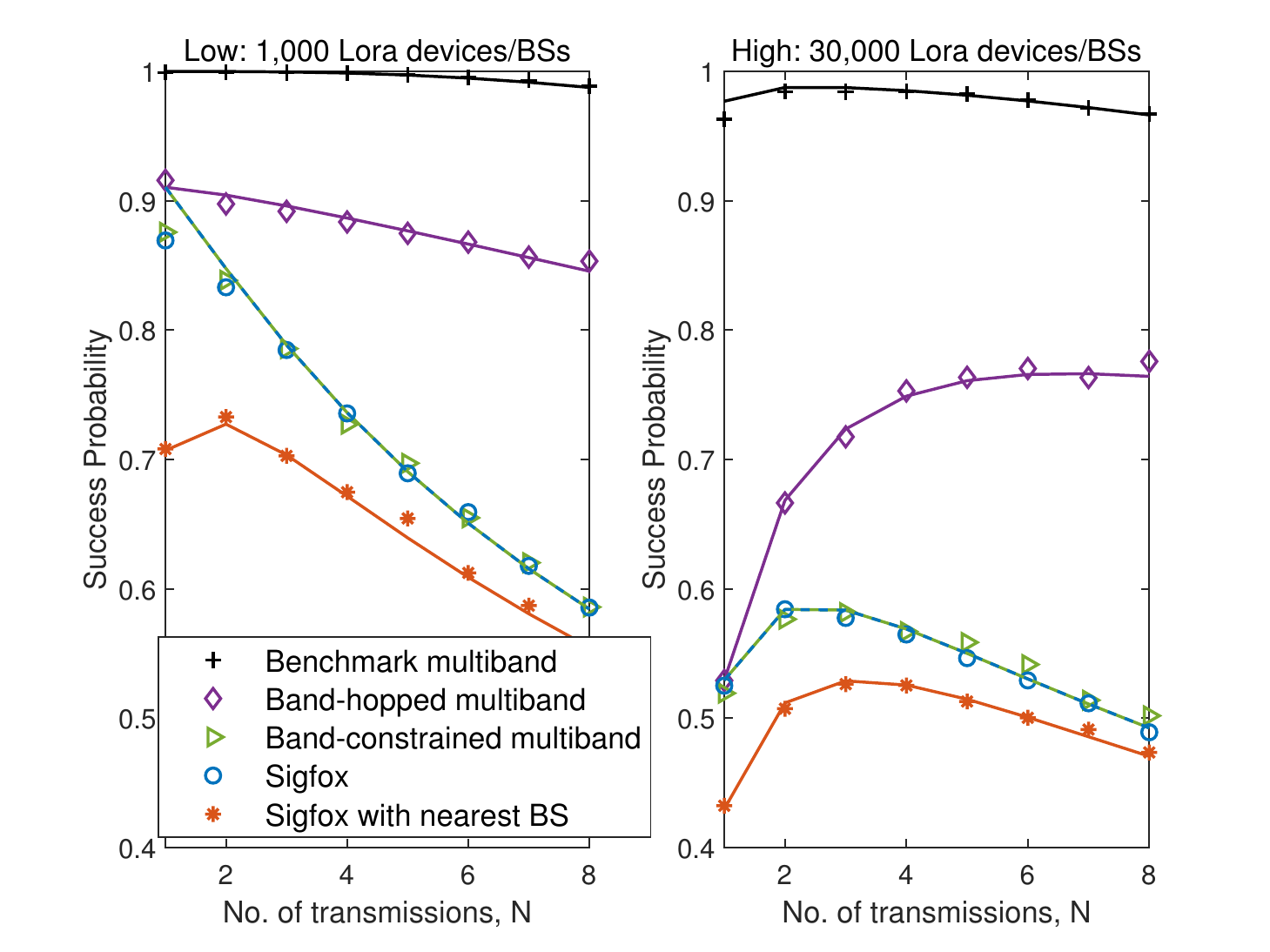}
			\caption{Variations of $N$ ($M = 5$ and Type-I incumbents)}
			\label{fig:Ps_vs_N}
		\end{subfigure}
		\caption{Success probability with variations of $M$ and $N$ ($\tau=0$dB).}
		\label{fig:Ps_vs_M_N}
	\end{figure}

	In Fig. \ref{fig:Ps_vs_M}, we show success probability with variations of the number of multiplexing bands, $M$. We also show the performance of single-band protocols for reference. It is observed that increasing the number of bands improves both the benchmark and proposed multiband protocols under Type-I, yet the latter has a saturating gain because as $M$ increases, the density of BSs listening to a particular band decreases, reducing the gain of geographical diversity. Under Type-II, we still observe gains with the benchmark scheme, but the band-hopped multiband becomes worse as $M$ becomes very high because not only geographical diversity is reduced, but also the new bands are already occupied by incumbents.
	
	In Fig. \ref{fig:Ps_vs_N}, we show the success probability with variations of $N$ and under different densities of Type-I incumbents. It is observed that for low-density interfering incumbents, a single transmission per packet is optimal for protocols with geographical diversity; however, the absence of this diversity under nearest BS association pushes the optimal number to two transmissions. Another interesting observation is that multiband access is less sensitive to the intra-network interference compared to single-band access, as allowing the device to have a larger pool of channels to select from reduces collisions. When the incumbents' density is high, then the optimal number under the benchmark multiband access and Sigfox is increased from one to two, which also asserts Corollary \ref{corollary:optimalN}, where it is shown that the optimal number is independent of $M$ if $B_{\operatorname{I}}<B\cdot M$. Further, in nearest BS association, the optimal number of transmissions is higher than that of Sigfox to compensate for the absence of geographical diversity. Likewise, the proposed multiband access protocol has also a higher optimal number of transmissions as we achieve a geographical diversity gain by sending different packets at different bands, where each packet is received by a different subset of BSs.

	%--------------------------------
	%Subsection: Transmission capacity comparison
	%-------------------------------- 
	\subsection{Transmission capacity comparison}
	
\begin{figure}[t!]
		\center
		\includegraphics[width=0.5\textwidth]{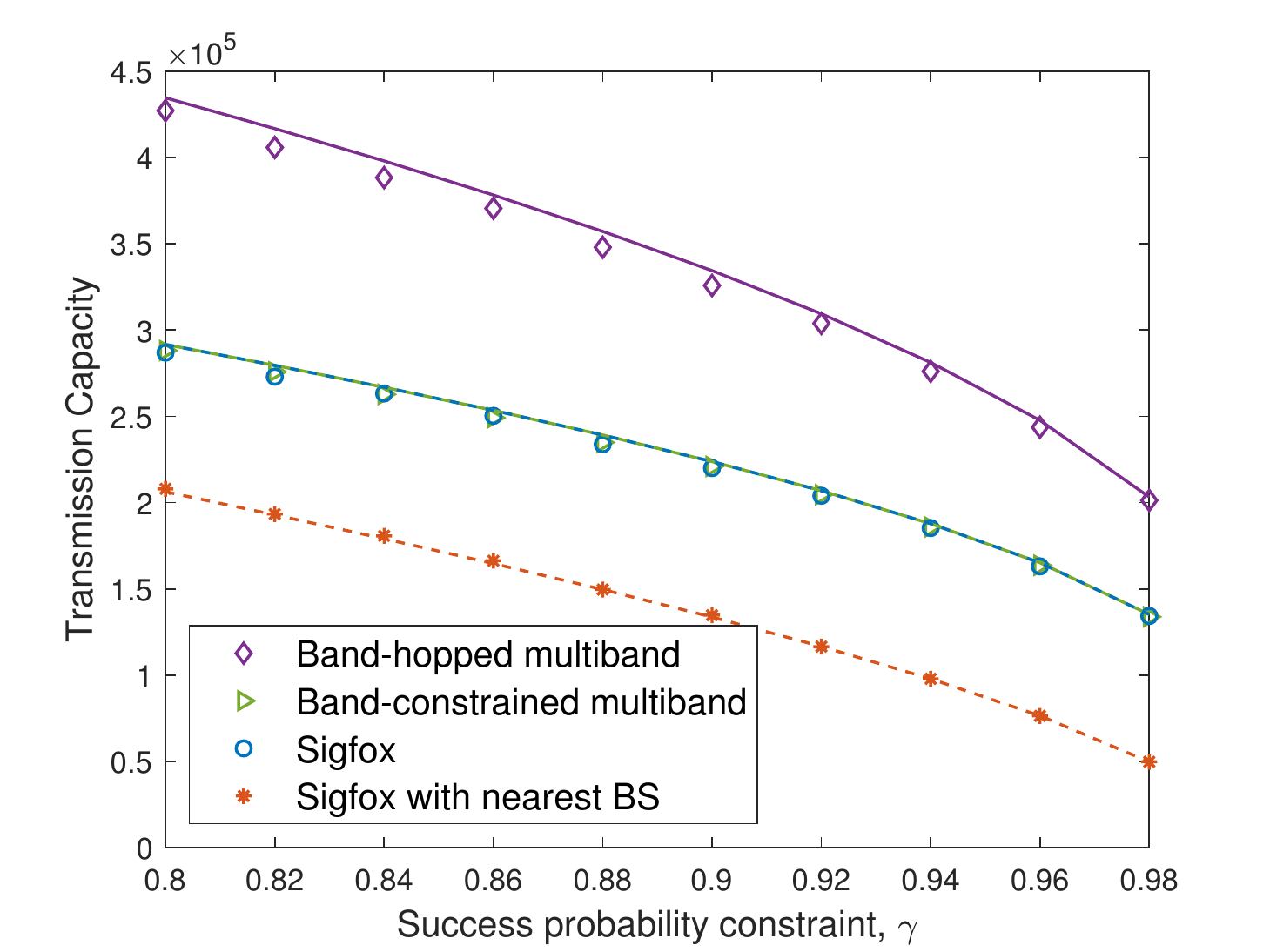}
		\caption{Transmission capacity for a given success probability constraint.}
		\label{fig:TC_vs_gamma}
	\end{figure}

	In Fig. \ref{fig:TC_vs_gamma}, we compare the different single-band and proposed multiband protocols in terms of the transmission capacity, where we assume $\tau=5$dB, $M=5$, and $N=3$. We assume the UNB network is deployed over an area of $25\times25$km$^2$ in the presence of Type-I network, and the average number of UNB BSs in that area is $25$. It is shown that UNB networks can provide coverage for a very large number of IoT devices. We observe that the proposed multiband protocol achieves the highest transmission capacity. For example, at a success probability constraint of 0.98, each UNB BS can support, on average, 8,000 IoT devices under the proposed protocol compared to only 2,000 IoT devices under Sigfox with nearest BS association.

	%-----------------------------------------------------------
	%Section: Conclusions
	%-----------------------------------------------------------
	\section{Conclusions}\label{sec:UNB_conclusion}
	
	An analytical framework has been developed to model and analyze UNB communications, an emerging paradigm that relies on ultra-narrowband signals to tackle the intra-network sharing among IoT devices and the inter-network sharing with incumbent networks. Several variants of UNB protocols are studied and compared in terms of the success probability and transmission capacity, where closed-form expressions are given for these metrics to identify fundamental limits and key performance trends under different network settings. 
	
	The analysis has shown that the geographical diversity, the number of signal repetitions, and the number of bands play key roles in the UNB network performance. First, the geographical diversity is essential for random access to support a massive number of IoT devices using only few BSs; otherwise, a high density of BSs is needed to enable massive access. Second, signal repetition is useful when the interference is dominated by incumbent networks, while a single transmission maximizes the success probability when the interference is dominated by UNB devices.  Third, using multiple bands reduces IoT collisions and interference with incumbents. However, to fully exploit these gains, all BSs are required to listen to all bands. When each BS is restricted to listen to one of the bands, then the diversity achieved by sending several repetitions is not sufficient as the density of UNB BSs listening to a given band is reduced. This elevates the need to optimize the BS-band selection policy. Alternatively, the IoT device can transmit each packet at a different band to better exploit the geographical diversity gain, even when the density of BSs per band is reduced. 
	
	Future work can include exploring BS-band assignment policies where coordination among BSs is exploited or machine learning is used to adapt theses assignments depending on the network status. Similarly, additional protocol enhancements, e.g., device-dependent number of repetitions, that utilize limited feedback from and to the UNB device can be considered.

	\appendices 
	%-----------------------------------------------------------
	%Appendix: Proof of theorems
	%-----------------------------------------------------------
	\section{Proof of Theorem \ref{th:Ps_nearest} and Theorem \ref{th:Ps} }\label{app:Ps}
	
	We first prove Theorem \ref{th:Ps_nearest}. We can simplify the expression in (\ref{eq:failProbR}) as follows 
	\begin{equation}
	\begin{array}{ll}
	\mathbb{Q}_j \\
	= \left(1-\mathbb{E}_{\tilde \Phi_{\IoT},\tilde\Phi_{\I},f_u,f_k} \left[e^{-\tau x_j^\alpha (\hat P_N +I_{\operatorname{UNB}} + I_{\operatorname{INC}})}\right]\right)^N\\
	= \left(1-e^{-\tau x_j^\alpha \hat P_N } \mathbb{E}_{\tilde \Phi_{\IoT},f_u}[e^{-\tau x_j^\alpha I_{\operatorname{UNB}}}] \mathbb{E}_{\tilde \Phi_{\I},f_k}[e^{-\tau x_j^\alpha I_{\operatorname{INC}}}]\right)^N.
	\end{array}
	\end{equation}
	Let  $s=\tau x_j^\alpha \hat P_{\I} $, then we have
	\begin{equation}
	\begin{aligned}
	\label{eq:INCPGFL}
	\mathbb{E}_{\tilde\Phi_{\I},f_k}[e^{-\tau x_j^\alpha I_{\operatorname{INC}}}] &= \mathbb{E}_{\tilde\Phi_{\I},f_k}\left[\exp\left(-s\textstyle \sum_{k\in\tilde\Phi_{\I}}  f_k y_{k,j}^{-\alpha}\right)\right]\\
	%&\stackrel{(a)}{=}  \mathbb{E}_{\tilde\Phi_{\I}}\left[\prod_{k\in\tilde\Phi_{\I}}\frac{1}{1+s y_{k,j}^{-\alpha}}\right]\\
	&\stackrel{(a)}{=}  \exp\left(- 2\pi \tilde{\lambda}_{\I}\textstyle \int_{0}^{\infty}  \big(y-\frac{y}{1+s y^{-\alpha}}\big)dy\right)\\
	&= \exp\left(-\pi \xi^{-1} (\tau\hat P_{\I})^\delta \tilde{\lambda}_{\I}x^2 \right),
	\end{aligned}
	\end{equation}
	where $(a)$ follows from using the characteristic function (CF) of $f\sim\exp(1)$ and the probability generating functional (PGFL) of the interfering incumbent HPPP. We note that the integral limit starts at zero because the interfering incumbent can be arbitrarily close to the BS. Likewise, let $\tilde s=\tau x_j^\alpha$, then we have $
	\mathbb{E}_{\tilde \Phi_{\IoT},f_u}[e^{-\tau x_j^\alpha I_{\operatorname{UNB}}}]= \exp\left(-\pi \xi^{-1}\tau^\delta  \tilde{\lambda}_{\IoT}x^2 \right)$.
	Plugging the above expressions into (\ref{eq:failProbR}) and using the binomial theorem, we get
	\begin{equation}
	\begin{aligned}
	\label{eq:Q_b_R}
	\mathbb{Q}_j
	&=\left(1-e^{-\pi \xi^{-1} \tau^\delta (\tilde{\lambda}_{\IoT}+\hat P_{\I}^\delta \tilde{\lambda}_{\I})x^2-\tau x_j^\alpha \hat P_N}\right)^N\\
	&= \sum_{k=0}^N \binom{N}{k} (-1)^k e^{-k\pi \xi^{-1} \tau^\delta (\tilde{\lambda}_{\IoT}+\hat P_{\I}^\delta \tilde{\lambda}_{\I})x^2-k\tau x_j^\alpha \hat P_N},
	\end{aligned}
	\end{equation}
	and hence
	\begin{equation}
	\label{eq:EQ_b_PNnearest}
	\begin{aligned}
	\mathbb{P}_s^{\operatorname{nearest}}&= 1-\mathbb{E}_{x}\textstyle\left[ \mathbb{Q}_j\right]\\
	&\stackrel{(a)}{=} 1- 2\pi \lambda_{\B} \sum_{k=0}^N \bigg\{\binom{N}{k} (-1)^k \\
	&\times\int_{0}^\infty  x e^{-\pi x^2 (\lambda_{\B}+k \xi^{-1} \tau^\delta ( \tilde{\lambda}_{\IoT}+\hat P_{\I}^\delta \tilde{\lambda}_{\I}))}dx\bigg\}\\
	&=1-\textstyle \sum_{k=0}^N \binom{N}{k} (-1)^k \left(1+\frac{k\xi^{-1} \tau^\delta ( \tilde{\lambda}_{\IoT}+\hat P_{\I}^\delta \tilde{\lambda}_{\I})}{\lambda_{\B}}\right)^{-1},
	\end{aligned}
	\end{equation}
	where $(a)$ follows from the distribution of the distance from the IoT device to its nearest BS, i.e., $f(x)=2\pi \lambda_{\B} x \exp(-\pi \lambda_{\B} x^2)$ \cite{Haenggi2012}, and the assumption $\hat P_N \rightarrow 0$.

	For Theorem \ref{th:Ps}, we can simplify (\ref{eq:PsGeneralNoAssociation}) by compute the following
	\begin{equation}
	\label{eq:EQ_b_R}
	\begin{array}{ll}
	\mathbb{E}_{\Phi_{\B}}\textstyle\left[\prod_{b\in\Phi_{\B}} \mathbb{Q}_b\right]\\
	=\exp\bigg(2\pi \lambda_{\B} \sum_{k=1}^N \bigg\{\binom{N}{k} (-1)^k  \\
	\times \int_{0}^\infty x e^{-\pi \xi^{-1} k\tau^\delta (\tilde{\lambda}_{\IoT}+\hat P_{\I}^\delta \tilde{\lambda}_{\I})x^2-k\tau x^\alpha \hat P_N} dx\bigg\}\bigg).
	\end{array}
	\end{equation}
	Plugging (\ref{eq:EQ_b_R}) in (\ref{eq:PsGeneralNoAssociation}), we get the exact success probability. We can further simplify the expression assuming an interference limited network, i.e., $\hat P_N\rightarrow0$. In this case, we have
	\begin{equation}
	\label{eq:integralNoNoise}
	\begin{aligned}
	\int_{0}^\infty x e^{-\pi\xi^{-1} \tau^\delta (k^\delta \tilde{\lambda}_{\IoT}+k\hat P_{\I}^\delta \tilde{\lambda}_{\I})x^2} dx = \frac{ \xi\tau^{-\delta}/(2\pi)}{k^\delta \tilde{\lambda}_{\IoT}+k\hat P_{\I}^\delta \tilde{\lambda}_{\I}}, 
	\end{aligned}
	\end{equation}
	and hence
	\begin{equation}
	\begin{array}{ll}
	\mathbb{E}_{\Phi_{\B}}\textstyle\left[\prod_{b\in\Phi_{\B}} \mathbb{Q}_b\right]\\
	=\exp\left(\xi \tau^{-\delta} \frac{\lambda_{\B}}{\tilde{\lambda}_{\IoT}+\hat P_{\I}^\delta\tilde{\lambda}_{\I}}\sum_{k=1}^N \binom{N}{k} \frac{(-1)^k}{k} \right).
	\end{array}
	\end{equation}
	Plugging (\ref{eq:integralNoNoise}) in (\ref{eq:EQ_b_R}) and then using (\ref{eq:PsGeneralNoAssociation}), we arrive at (\ref{eq:Ps}).
	
	%-----------------------------------------------------------
	%Appendix: Solution of the optimization problem 
	%-----------------------------------------------------------
	\section{Solution of the optimization problem in (\ref{eq:SMBopt})}\label{app:dual}
	Let $\nu$ be the Lagrange multiplier for the constraint $\sum_{m=1}^Mp_m=1$ and $\mu_m$ be the multiplier for the constraint $p_m\geq 0$. The Lagrangian is then given as
	\begin{equation}
	L(p_m,\mu_m,\nu) = \sum_{m=1}^M e^{-c_mp_m} + p_m(\nu-\mu_m) - \nu.
	\end{equation}
	From the Karush–Kuhn–Tucker (KKT) conditions, we have 
	\begin{equation}
	\label{eq:kkt}
	-c_m e^{-c_mp_m^\star} + \nu^\star - \mu_m^\star = 0 \Longleftrightarrow \nu^\star \geq c_m e^{-c_mp_m^\star}
	\end{equation}
	since $\mu_m^\star \geq0$.
	
	In addition, from the complementary slackness condition, i.e., $p_m^\star \mu_m^\star =0$, we have $p_m^\star(\nu^\star-c_m e^{-c_mp_m^\star}) = 0$. We make the following observations. First, if $\nu^\star < c_m$, then (\ref{eq:kkt}) implies that  $p_m^\star > 0$, and from the complementary slackness condition, we have $\nu^\star = c_m e^{-c_mp_m^\star}$ or $p_m^\star = \frac{1}{c_m}\ln\left(\frac{c_m}{\nu^\star}\right)$. If $\nu^\star \geq c_m$, then we must have $p_m^\star = 0$; otherwise, $\nu^\star \geq c_m >  c_m e^{-c_mp_m^\star}$, which violates the complementary slackness condition. Thus, we have proved that $
	p_m^\star = \max\left\{0,\frac{1}{c_m}\ln\left(\frac{c_m}{\nu^\star}\right)\right\}
	$.
	Finally, from the KKT condition $\sum_{m=1}^M p_m^\star = 1$, we get  $
	\sum_{m=1}^{M} \max\left\{0,\frac{1}{c_m}\ln\left(\frac{c_m}{\nu^\star}\right)\right\}  = 1$.

	%---------------------------------------------------------------------------------
	%                         References
	%---------------------------------------------------------------------------------
	\bibliographystyle{IEEEtran}
	\bibliography{C:/Users/ghait/Dropbox/References/IEEEabrv,C:/Users/ghait/Dropbox/References/References}
	
\end{document}